\documentclass[manuscript]{aastex}
\usepackage{color}
\usepackage{leftidx}

\usepackage[tight]{subfigure}
\usepackage[colorlinks]{hyperref}
\hypersetup{ colorlinks, linkcolor=red, citecolor=blue }

\slugcomment{accepted for publication in ApJ}

\shorttitle{The impact of SN Ia explosions on their He companions}
\shortauthors{Z.-W. Liu et al.}

\begin{document}


\title{The impact of Type Ia supernova explosions on helium companions in the Chandrasekhar-mass explosion scenario}

\author{Zheng-Wei Liu\altaffilmark{1,2,3,4}, R. Pakmor\altaffilmark{5}, I.~R. Seitenzahl\altaffilmark{4,6}, W. Hillebrandt\altaffilmark{4}, M. Kromer\altaffilmark{4}, F.~K. R\"opke\altaffilmark{6}, P. Edelmann\altaffilmark{4}, S. Taubenberger\altaffilmark{4}, K. Maeda\altaffilmark{7}, B. Wang\altaffilmark{1,2} and Z. W. Han\altaffilmark{1,2}}

\email{zwliu@ynao.ac.cn}

\altaffiltext{1}{Yunnan Observatories, Chinese Academy of Sciences, Kunming 650011, P.R. China}
\altaffiltext{2}{Key Laboratory for the Structure and Evolution of Celestial Objects, Chinese Academy of Sciences, Kunming 650011, P.R. China}
\altaffiltext{3}{University of Chinese Academy of Sciences, Beijing 100049, P.R. China}
\altaffiltext{4}{Max-Planck-Institut f\"ur Astrophysik, Karl-Schwarzschild-Str. 1, 85741 Garching, Germany}
\altaffiltext{5}{Heidelberger Institut f\"ur Theoretische Studien, Schloss-Wolfsbrunnenweg 35, 69118 Heidelberg, Germany}
\altaffiltext{6}{Institut f\"ur Theoretische Physik und Astrophysik, Universit\"at W\"urzburg, Am Hubland, 97074 W\"urzburg, Germany}
\altaffiltext{7}{Kavli Institute for the Physics and Mathematics of the Universe (Kavli-IPMU), Todai Institutes for Advanced Study (TODIAS),
University of Tokyo, 5-1-5 Kashiwanoha, Kashiwa, Chiba 277-8583, Japan}

\begin{abstract}
  In the version of the single-degenerate scenario of Type Ia
  supernovae (SNe~Ia) studied here, a carbon--oxygen white dwarf
  explodes close to the Chandrasekhar limit after accreting material
  from a non-degenerate helium (He) companion star.  In the present study, we employ
  the {\sc Stellar GADGET} code to perform three-dimensional
  hydrodynamical simulations of the interaction of the SN~Ia ejecta
  with the He companion star taking into account its orbital motion
  and spin.  It is found that only $2\%-5\%$ of the initial companion
  mass are stripped off from the outer layers of He companion stars
  due to the SN impact.  The dependence of the unbound mass (or the
  kick velocity) on the orbital separation can be fitted in good
  approximation by a power law for a given companion model. After the
  SN impact, the outer layers of a He donor star are significantly
  enriched with heavy elements from the low-expansion-velocity tail of
  SN~Ia ejecta.  The total mass of accumulated SN-ejecta material on
  the companion surface reaches about $\gtrsim
  10^{-3}\,\mathrm{M}_{\odot}$ for different companion models. This
  enrichment with heavy elements provides a potential way to
  observationally identify the surviving companion star in SN
  remnants. Finally, by artificially adjusting the explosion energy of
  the W7 explosion model, we find that the total accumulation of SN
  ejecta on the companion surface is also dependent on the explosion
  energy with a power law relation in good approximation.
\end{abstract}


\section{INTRODUCTION}
 \label{sec:introduction}

 Type Ia supernovae (SNe~Ia) are instrumental as distance indicators
 on a cosmic scale to determine the expansion history of the Universe
 \citep{Ries98, Schm98, Perl99}. However, neither observational nor
 theoretical approaches have been able to identify the nature of
 SN~Ia progenitors and details of the explosion mechanism remain
 unclear (see \citealt{Hill00, Hill13} for a review).  Recently, the
 nearby SN 2011fe \citep{Nuge11, Li11} has been used as an important
 test case to constrain SN~Ia explosion scenarios \citep{Roep12} since
 it can be observed in unprecedented detail. However, additional
 investigations are still required to put more constraints on SN~Ia
 explosions.
 
 It is widely believed that SNe~Ia originate from thermonuclear
 explosions of carbon--oxygen (CO) white dwarfs (WDs) in binary
 systems. Depending on the nature of the companion star, the most
 favored progenitor models of SNe~Ia are classified into two general
 categories, the \emph{``single-degenerate''} (SD) scenario
 \citep{Whel73, Nomo82} and the \emph{``double-degenerate''} (DD)
 scenario \citep{Whel73, Iben84, Webb84}.  In the DD scenario, two CO
 WDs spiral in and merge due to gravitational wave radiation, causing
 a thermonuclear explosion of the merged system.  Recently, some
 observational and hydrodynamical studies support the viability of DD
 models as the progenitors of SNe~Ia (see, e.g., \citealt{Li11,
   Nuge11, Chom12, Hore12, Bloo12, Scha12, Pakm10, Pakm11, Pakm12b,
   Pakm13}).  In contrast, previous simulations suggested that the DD
 scenario likely leads to an accretion-induced collapse rather than a
 SN~Ia \citep{Nomo85, Timm94}.

\begin{figure}
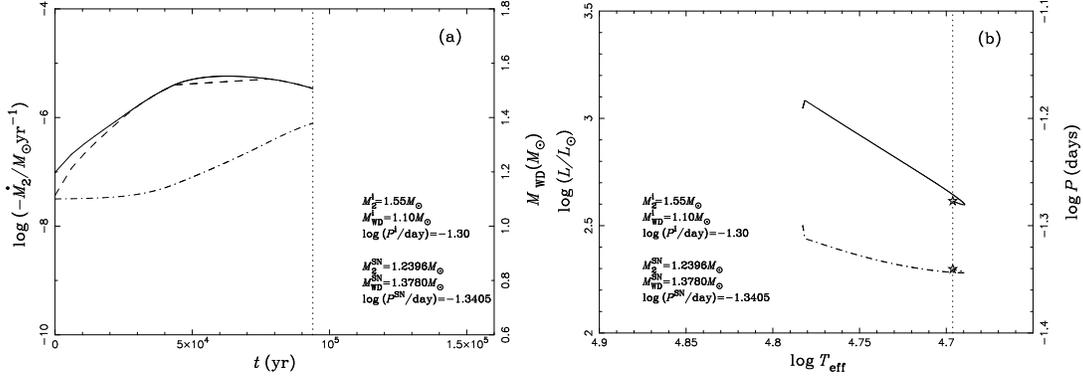

  \begin{center}
    \subfigure
    {\includegraphics[width=0.3\textwidth, angle=270]{f1a.eps}}
    \subfigure
    {\includegraphics[width=0.3\textwidth, angle=270]{f1b.eps}}
    \caption{ {\it Panel (a)\/}: the solid, dashed and dash-dotted curves
      show the mass transfer rate from the secondary, $\dot M_{2}$,
      the mass growth rate of the CO WD, $\dot M_{\mathrm{CO}}$, and
      the mass of the CO WD, $M_{\mathrm{WD}}$, varying with time,
      respectively. {\it Panel (b)\/}: the evolutionary track of the He
      donor star is shown as a solid curve and the evolution of
      orbital period is shown as a dash-dotted curve. Note that the He
      companion is still a MS star at the moment of the SN explosion.}
\label{Fig:hrd1}
  \end{center}
\end{figure}

\begin{figure}
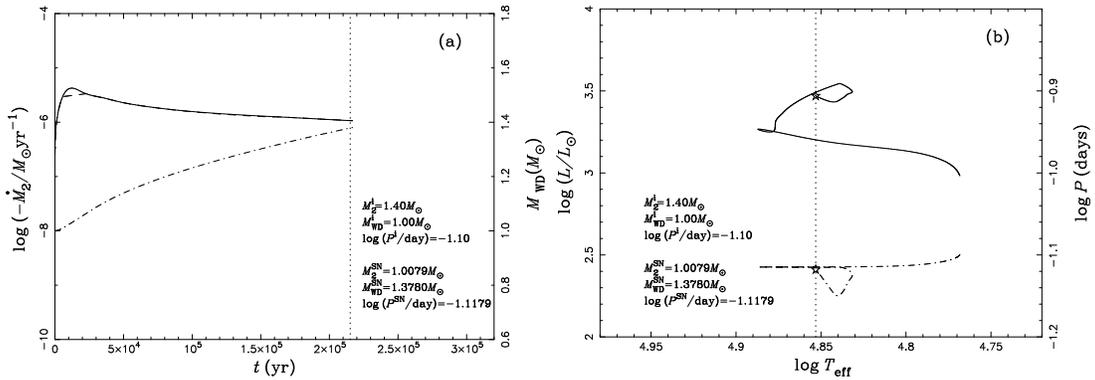

  \begin{center}
    \subfigure
    {\includegraphics[width=0.3\textwidth, angle=270]{f2a.eps}}
    \subfigure
    {\includegraphics[width=0.3\textwidth, angle=270]{f2b.eps}}
    \caption{Same as Figure~\ref{Fig:hrd1}, but for the He companion
      star that slightly evolves to the subgiant (SG) phase at the
      moment of SN explosion.}
\label{Fig:hrd2}
  \end{center}
\end{figure}

In the SD scenario, a CO WD increases its mass by accreting material
from a non-degenerate companion star (a slightly evolved main sequence
star [MS], a red giant [RG] or a He star [HE]) to approach the
critical explosion mass (just below the Chandrasekhar limit
$M_{\rm{Ch}}\sim1.44\,\mathrm{M}_{\odot}$) to ignite a SN~Ia
explosion. There is evidence from observations supporting that the
progenitors of some SNe~Ia come from the SD channel. For example, the
pre-SN circumstellar matter has been detected, and the features of the
interaction of the SN explosion with circumstellar matter are seen in
observations (see, e.g., \citealt{Pata07, Ster11, Fole12, Dild12,
  Shen13}). However, only a fairly narrow range of accretion rates is
allowed in order to avoid nova explosions in the SD case, making it
difficult to explain the observed nearby SN~Ia rate\citep{Nomo82,
  Nomo07, Ruit09, Wang10b}.

In recent years, the WD+MS and WD+RG progenitor models have been
invoked to explain the observed long-delay-time ($\gtrsim
1\,\mathrm{Gyr}$) population of SNe~Ia (see, e.g., \citealt{Ruit09,
  Wang10a, Wang10b, Maoz10a, Maoz10b, Maoz12}).  Numerically, the
impact of a SN~Ia explosion on a MS-like or a RG companion star has
been studied with hydrodynamical simulations by several authors (see,
e.g., \citealt{Mari00, Pakm08, Pan10, Pan12, Liu12, Liu13}).  They found that
$\sim0.03$--$0.15\,\mathrm{M}_{\odot}$ H-rich material can be stripped
off from the surface of MS companion stars. For RG companions it is
believed that the entire envelope is stripped off by the SN impact.
This high stripped mass is far larger than the most stringent upper
limit of $0.01\,M_{\odot}$ which \citet{Leon07} derived from the
non-detection of $\mathrm{H}_{\alpha}$ emission in late time spectra
(see also \citealt{Shap12}). So far, in fact, no direct observation
shows the signature of stripped H-rich material, which seriously
challenges SD progenitor scenario.

In the spin-up/spin-down model for SNe Ia\footnote{In the SD scenario,
  a WD accretes and retains companion matter that carries angular
  momentum. As a consequence the WD spins with a short period which
  leads to an increase of the critical explosion mass. If the critical
  mass is higher than the actual mass of the WD, the SN explosion can
  only occur after the WD increases the spin period with a specific
  spin-down timescale (see \citealt{Di11, Just11}).}, however, the
donor star might shrink significantly because it exhausts the H-rich
envelope before the explosion sets in after a spin-down phase of
$>10^{5}\,\mathrm{yrs}$. Thus, the donor star could be too dim to
detect by the time of explosion and much smaller than its Roche lobe
(see \citealt{Di11, Just11}).  This may reduce the possibility of the
detection of H lines in SN~Ia nebular spectra and possibly provides an
explanation for the apparent lack of a `left-over' star in LMC SN
remnant $\mathrm{SNR\,0609-67.5}$ \citep{Di12}.

In the so-called WD+HE channel a CO WD accretes material from a He
companion star.  This may initiate a thermonuclear explosion of the
WD.  At present, two possible explosion models are frequently
discussed: the sub-$M_{\rm{Ch}}$ scenario \citep{Woos94, Fink07, Fink10,
  Woos11} and the $M_{\rm{Ch}}$ scenario \citep{Wang09a, Wang09b}. 
 In this paper, however, we only focus on the WD+HE
$M_{\rm{Ch}}$ model.  With a binary population synthesis (BPS)
approach, \citet{Wang09a} (hereafter WMCH09) comprehensively and
systematically investigated WD+HE $M_{\rm{Ch}}$ systems and showed
that this channel can explain SNe~Ia with short-delay times
($\lesssim10^{8}\,\mathrm{yrs}$), which is consistent with recent
observational implications of young populations of SN~Ia progenitors
\citep{Wang09a, Wang09b}.

\begin{table*} \renewcommand{\arraystretch}{1.0}\scriptsize
\begin{center}
\caption{Results of SPH impact simulations. \label{table:1}}
\begin{tabular}{lrrccccccccc}
\tableline\tableline
Model\tablenotemark{a}& $v_{\mathrm{orb}}$ & $v_{\mathrm{spin}}$ & $R_2$& $A$  &$M_{\mathrm{unbound}}$ 
&$v_{\mathrm{kick}}$\ \ \  &\multicolumn{1}{c}{ $\delta M_{\mathrm{tot}}$\tablenotemark{b}}& \multicolumn{1}{c}{ $\delta M_{\mathrm{Fe}}$\tablenotemark{b}} & \multicolumn{1}{c}{ $\delta M_{\mathrm{Ni}}$\tablenotemark{b}} & $M_{\mathrm{Ni}}/M_{\mathrm{He}}$ \tablenotemark{c} &  $M_{\mathrm{Fe}}/M_{\mathrm{He}}$ \tablenotemark{c}\\
& \multicolumn{2}{c}{$(\mathrm{km\,s^{-1}})$} & \multicolumn{2}{c}{$(\mathrm{10^{10}\,cm})$} &\multicolumn{1}{c}{$(\mathrm{M}_{\odot})$} & $(\mathrm{km\,s^{-1}})$ &\multicolumn{3}{c}{$(10^{-3}\,\mathrm{M}_{\odot})$} & $(10^{-3})$ & $(10^{-3})$\\ 
\tableline
W7\_He01 & -&-             & 1.91 & 5.16 &0.027 & 66.39 
& 5.22 \ \ \  & 3.52 & 1.59 & 1.63  & 3.62\\
W7\_He02 & -&-             & 2.48 & 7.04 &0.056 & 58.75 
& 3.12 \ \ \  & 2.16 & 0.88 & 3.54 & 8.72\\
W7\_He01\_r & 432  & 301 & 1.91 & 5.16 &0.028 & 66.94 
& 5.37 \ \ \  & 3.49 & 1.81 & 1.85  & 3.57\\
W7\_He02\_r & 387  & 237 & 2.48 & 7.04 &0.057 & 59.74 
& 3.14 \ \ \  & 2.02 & 0.96 & 3.86 & 8.11\\
W708\_He01\_r & 432  & 301 & 1.91 & 5.16 &0.019 & 39.93
& 12.06 \ \ \ & 7.49 & 4.44 & 4.52 & 7.61\\
W710\_He01\_r & 432  & 301 & 1.91 & 5.16 &0.024 & 52.53 
& 8.30  \ \ \ & 5.29 & 2.91 & 2.97 & 5.39\\
W714\_He01\_r & 432  & 301 &1.91 & 5.16 &0.033 & 75.89 
& 3.97 \ \ \  & 2.57 & 1.38 & 1.38  & 2.64\\
W716\_He01\_r & 432  & 301 &1.91 & 5.16 & 0.037 & 86.08
& 2.71 \ \ \  & 1.80 & 0.87 & 0.90  & 1.86\\
\tableline
\end{tabular}
\tablecomments{\footnotesize \\ $^{\mathrm{a}}$ ``W7'' corresponds to the W7
    explosion model \citep{Nomo84, Maed10}.  ``W708'', ``W710'',
    ``W714'' and ``W716'' present W7-like models that are produced by
    adjusting the original W7 model with different explosion energies
    ($0.8, 1.0, 1.4$ and $1.6 \times 10^{51}\,\mathrm{erg}$). Note
    that all parameters but the SN energy are kept constant with the
    values of the original model (see also \citealt{Pakm08}). ``He01''
    and ``He02'' are two He companion models. ``r'' means that the
    orbital motion and spin of the He companion are included. \\
    $^{\mathrm{b}}$ $\delta M_{\mathrm{tot}}$, $\delta
    M_{\mathrm{Fe}}$ and $\delta M_{\mathrm{Ni}}$ are the total
    contamination, the accreted Fe and Ni mass at the end of the
    simulations ($\gtrsim 2000\,\mathrm{s}$ after the explosion),
    respectively. \\ 
    $^{\mathrm{c}}$ The ratio of bound Ni and
    Fe masses (without decay) to the He masses of a surviving
    star. Please note that the initial metallicity of the He star is
    not included.  }
\end{center}
\end{table*}

Recently, \citet{Pan10, Pan12} investigated the impact of SN~Ia ejecta
on a He companion star including the rotation of the He star by using
Eulerian hydrodynamics simulations with the {\sc FLASH} code.  They
found the He companion star could be contaminated by the SN~Ia ejecta
in its outer envelope after the impact, and the nickel contamination
is ${\sim}10^{-4}\,M_{\odot}$ \citep{Pan10, Pan12}.  This might help
to identify surviving companion stars in the remnants of historical
SNe~Ia even a long time after the explosion. In their simulations,
however, the He star companion models were constructed by artificially
adopting a constant mass-loss rate to mimic the detailed binary
evolutionary models of WMCH09.

In this work, we update the He companion star models with
one-dimensional (1D) consistent binary evolution calculations.  Then,
we perform hydrodynamics simulations of the interaction of SN~Ia
ejecta with He companion stars. To this end we use the
three-dimensional (3D) smoothed particle hydrodynamics (SPH) code {\sc
  Stellar GADGET}.  In Section~\ref{sec:codes}, the code and the
initial setup are introduced. The results of the SPH impact
simulations are discussed on the basis of two consistent He companion
star models in Section~\ref{sec:simulations}.  All numerical results
are presented in Section~\ref{sec:discussion}. Finally, we summarize
our results and conclude in Section~\ref{sec:conclusions}.


\section{CODES AND INITIAL MODELS}
  \label{sec:codes}

  We use Eggleton's stellar evolution code \citep{Eggl71, Eggl72,
    Eggl73} to follow the detailed binary evolution of WD+HE
  progenitor systems.  The Roche lobe overflow (RLOF) is treated in
  the code as described by \citet{Han04}.  In this work, we only focus
  on $M_{\rm{Ch}}$ explosions of accreting WDs.  The influence of
  rotation on the He-accreting WDs is not considered in the stellar
  evolution calculations.  Our basic input physics and initial setup in
  the code are the same as those in WMCH09.  The He companion star is
  evolved without enhanced mixing, i.e., the convective overshooting
  parameter, $\delta_{\mathrm{ov}} = 0$ (see \citealt{Dewi02}).
  Initial He star models are set up with a He abundance of $\rm{Y =
    0.98}$ and a metallicity of $\rm{Z = 0.02}$.  In addition, orbital
  angular momentum loss due to gravitational wave radiation (GWR) is
  included by adopting a standard formula presented by \citet{Land71}:

   \begin{equation}
    \label{eq:1}
      \frac{{\mathrm{d\ ln}} J_{\mathrm{GR}}}{{\mathrm{d}}t} = - \frac{32G^3}{5c^5} \frac{M_{\mathrm{WD}}M_{\mathrm{2}}(M_{\mathrm{WD}}+M_{\mathrm{2}})}{A^4},
   \end{equation}

   where $G,c, M_{\mathrm{WD}}$ and $M_{\mathrm{2}}$ are the
   gravitational constant, vacuum speed-of-light, mass of the
   accreting WD and mass of the He companion star, respectively.

\begin{figure}
\centering
\includegraphics[width=0.45\textwidth, angle=360]{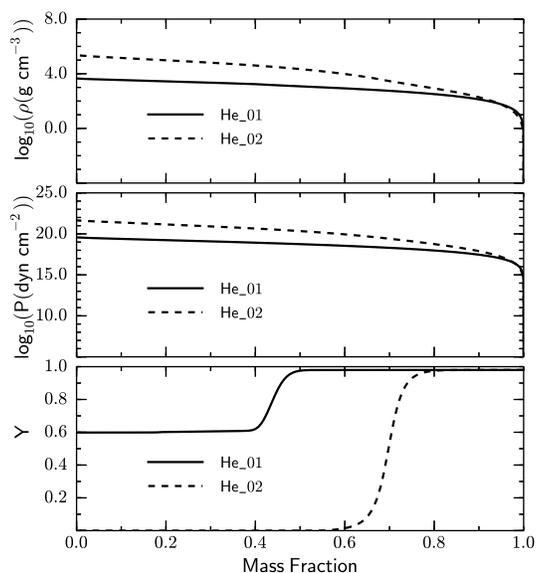}
\caption{Profiles of the density $\rho$, pressure $P$, and helium abundance $Y$ 
      as a function of enclosed mass $m$ at the moment of the SN explosion for 
      the He01 model (solid lines) and He02 model (dashed lines).}
\label{Fig:structure}
\end{figure}

\begin{figure}
\centering
\includegraphics[width=0.45\textwidth, angle=360]{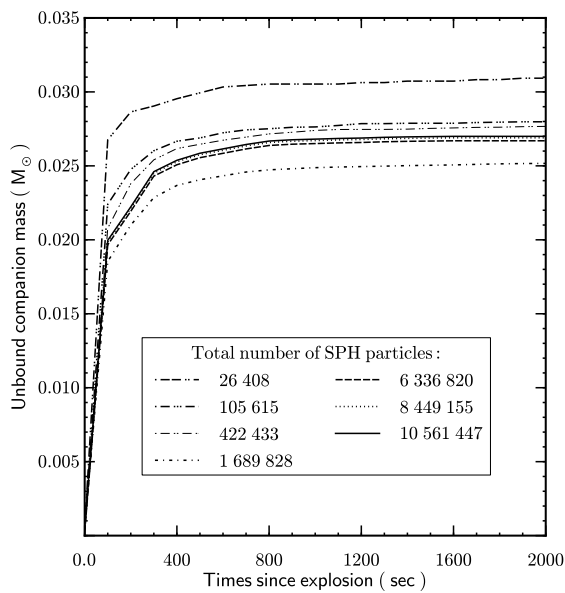}
\caption{ Unbound companion mass vs. time since explosion in W7\_He01
  model for different resolutions (${\sim}10^4-10^7$ SPH particles in
  the simulations).}
\label{Fig:resolution}
\end{figure}

\begin{figure*}
  \begin{center}
  \includegraphics[width=0.85\textwidth, angle=0]{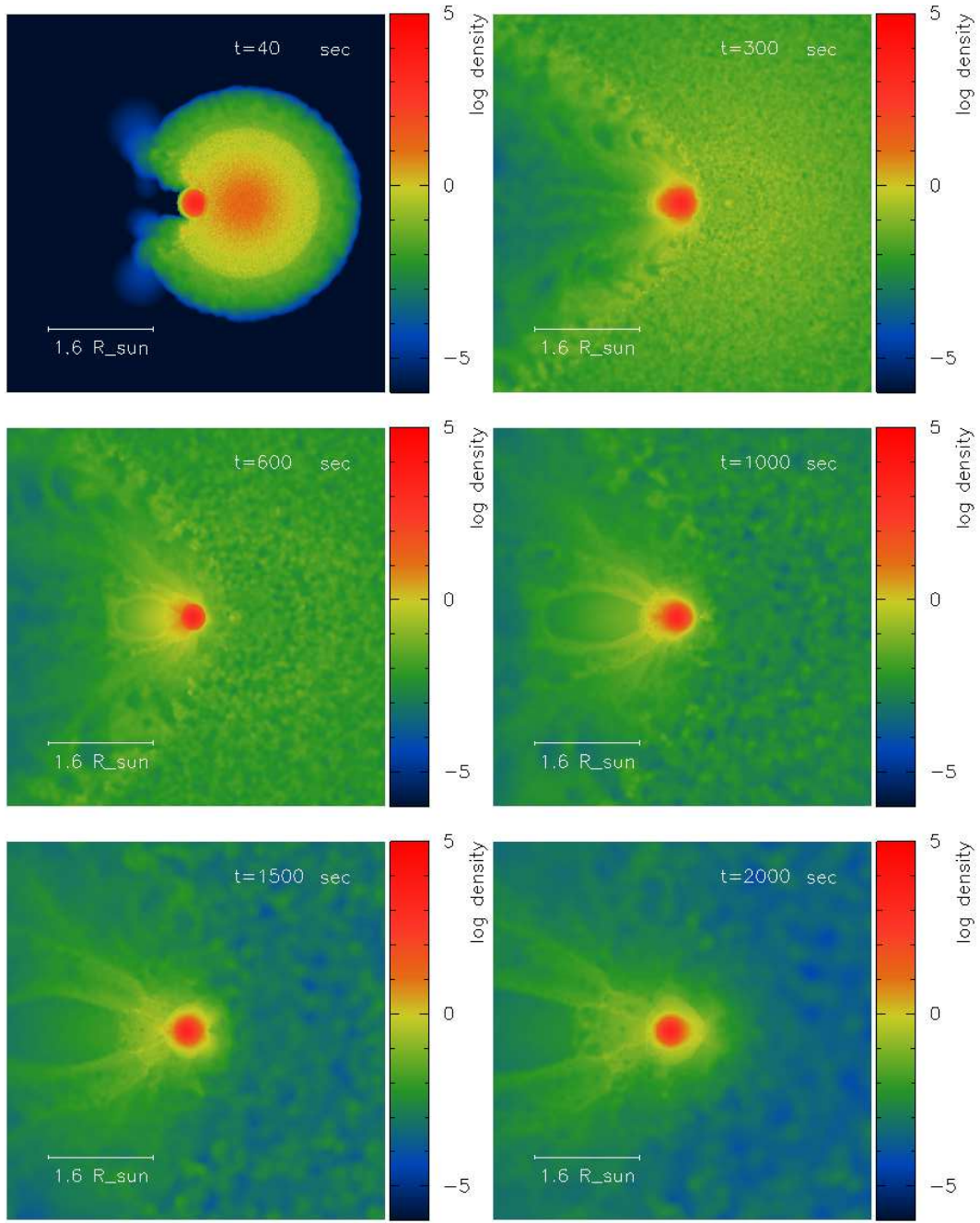}
  \caption{Temporal evolution of the density structure of SN and
    companion material in impact simulations with W7\_He01 model. The
    color scale indicates the logarithm to base 10 of density in $\rm{g\,cm^{-3}}$. The
    plots are made using the freely available {\sc SPLASH} code
    \citep{Pric07}.}
    \label{Fig:rho}
  \end{center}
\end{figure*}

\begin{figure*}
  \begin{center}
  \includegraphics[width=0.85\textwidth, angle=0]{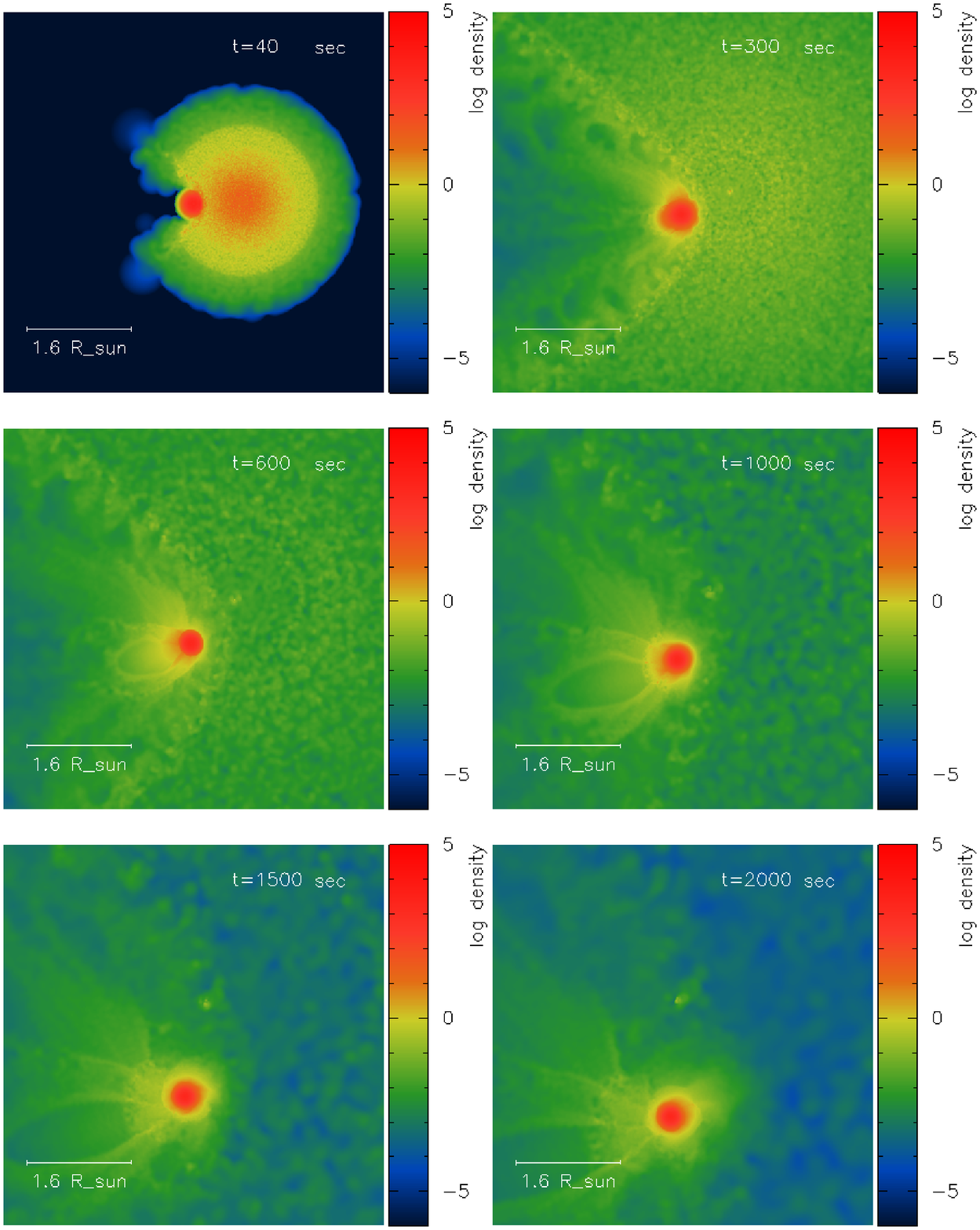}
  \caption{Same as Figure~\ref{Fig:rho}, but for the W7\_He01\_r model
    (which includes the orbital motion and spin of He companion). The
    color scale indicates the logarithm to base 10 of density in $\rm{g\,cm^{-3}}$. The
    plots are made using the freely available {\sc SPLASH} code
    \citep{Pric07} }
    \label{Fig:rho_rot}
  \end{center}
\end{figure*}

We start to trace the binary evolution when the WD+HE binary system is
formed.  The mass transfer occurs through RLOF once the He donor star
fills its Roche lobe. Here, we do not solve the stellar structure
equations for the WD star when the structures of the companion stars
are constructed.  Instead, we used the optically thick wind model of
\citet{Hach96, Hach99} and adopt the prescription of \citet{Kato04}
for the mass accumulation efficiency of He-shell flashes onto the 
WD primary. If the mass transfer
rate, $\dot{M}_{\mathrm{2}}$, is above a critical value,
$\dot{M}_{\mathrm{cr}}$, we assume that He burns steadily on the
surface of the WD and that the He-rich material is converted into
carbon and oxygen at the rate $\dot{M}_{\mathrm{cr}}$, while the
unprocessed matter is assumed to be lost from the system as an
optically thick wind at a mass-loss rate $\dot{M}_{\mathrm{wind}}=
\left | \dot{M}_{2} \right | - \dot{M}_{\mathrm{cr}}$.  The critical
mass-accretion rate is \citep{Nomo82}

  \begin{equation}
    \label{eq:2}
     \dot M_{\mathrm{cr}} = 7.2 \times 10^{-6} (M_{\mathrm{WD}}/\mathrm{M_{\odot}}-0.6)\ \mathrm{\mathrm{M}_{\odot}\,yr^{-1}},
  \end{equation}

  When $\left |\dot M_{\mathrm{2}} \right |$ is smaller than $\dot
  M_{\mathrm{cr}}$, the following assumptions have been adopted:

  \begin{enumerate}
  \item If $\dot{M}_{\mathrm{st}} \leqslant |\dot{M}_{\mathrm{2}}|
    \leqslant \dot{M}_{\mathrm{cr}}$, it is assumed that there is no
    mass loss and that He-shell burning is steady, where
    $\dot{M}_{\mathrm{st}}$ is the minimum accretion-rate of stable
    He-shell burning from \citet{Kato04}.
  \item If $\dot M_{\mathrm{low}} \leqslant \left |\dot M_{\mathrm{2}}
    \right | < \dot M_{\mathrm{st}}$, He-shell burning is unstable,
    He-shell flashes occur and a part of the envelope mass is assumed
    to be blown off from the surface of the WD. Here, $\dot
    M_{\mathrm{low}} = 4.0 \times
    10^{-8}\,\mathrm{M}_{\odot}\,\mathrm{yr^{-1}}$ is the minimum
    accretion-rate of weak He-shell flashes \citep{Woos86}.
  \item If $\left |\dot M_{\mathrm{2}} \right | < \dot
    M_{\mathrm{low}}$, He-shell flashes are so strong that no mass can
    be accumulated by the WD (i.e., the mass-growth rate of the WD is
    zero).
    
  \end{enumerate}

  Finally, two He companion star models based on detailed binary
  evolution calculations are chosen as representative examples to
  perform SPH impact simulations.  They are named with ``He01'' and
  ``He02''. The typical binary evolution calculations of these two
  models are shown in Figure~\ref{Fig:hrd1} (He01 model) and
  Figure~\ref{Fig:hrd2} (He02 model).  In the He01 model (companion mass $M_{2}^{\rm{SN}}=1.2396\,\rm{M_{\odot}}$, orbital
  separation $A=5.16 \times 10^{10}\,\mathrm{cm}$, companion radius
  $R_2=1.91 \times 10^{10}\,\mathrm{cm}$) the companion star remains
  to be a He MS star (central He burning), whereas in the He02 model
  ($M_{2}^{\rm{SN}}=1.0079\,\rm{M_{\odot}}$, $A=7.04 \times 10^{10}\,\mathrm{cm}$, $R_2=2.48 \times
  10^{10}\,\mathrm{cm}$) it has evolved slightly into the subgiant
  phase (central He exhausted) at the onset of the SN~Ia explosion.
  The structure profiles of two companion stars (the He01 and He02 model) at the moment of SN Ia  
  explosion are shown in Figure~\ref{Fig:structure}.

  The hydrodynamical simulations of the impact of SN~Ia ejecta on the
  He companion star are performed with the 3D SPH code {\sc Stellar
    GADGET} \citep{Pakm12a, Spri05}.  In our simulation, we use the same
  method as \citet{Liu12} to map the 1D profiles of density and
  internal energy of a 1D companion star model to a particle
  distribution suitable for the SPH code. To reduce numerical noise introduced by the mapping, the
  SPH model of each companion star is relaxed for several dynamical
  timescales ($1.0 \times 10^4\,\mathrm{s}$) before we start the
  actual impact simulations. If the relaxation succeeds, the
  velocities of the particles stay close to zero. Otherwise, we reject
  the SPH model, and repeat the relaxation after adjusting the
  relaxation parameters \citep{Pakm12a}.

  The SN explosion is represented by the W7 model of \citet{Nomo84,
    Maed10}. This model has been shown to provide a good fit to the
  observational light curves of SNe~Ia \citep{Lent01}.  Its total explosion energy is
  1.23 $\times$ 10$^{51}\,\mathrm{erg}$, the average velocity of the
  ejecta $10^4\,\mathrm{km\,s}^{-1}$. Based on the 1D W7 model of \citet{Nomo84}, SPH particles are placed
  randomly in shells to reproduce the mass (density) profile and gain the radial velocities 
  they should have at their positions (all particles have the same mass). The composition of a particle is then set to
  the values of the initial 1D model at a radius equal to the radial coordinate of the particle. In our hydrodynamical
  simulations, the impact of the SN~Ia ejecta on the companion is
  simulated for $\mathrm{\geq2000\,s}$ taking into account the orbital
  motion and spin of the He companion star. Here, we assume that the rotation
  of the companion star is in phase-locked to its orbital motion. Moreover, we set the $x-y$
  plane as the orbital plane of the binary system with an assumption
  of a circular orbit.  The $z-$axis is chosen as the rotation axes,
  and, when the spin of the companion star is included, the positive
  $z-$axis is the direction of the angular momentum.

\section{SIMULATIONS}
\label{sec:simulations}

  \subsection{Resolution Test}
  \label{sec:convergence}

  We use the W7\_He01 model (see Table~\ref{table:1}) as a typical
  case to perform a convergence test to check the sensitivity of
  unbound mass to different resolutions. By adopting a fixed orbital
  separation ($A = 5.16\times10^{10}\,\rm{cm}$), the resolutions are
  set up with different number of total SPH particles ranging from
  $2.64\times10^{4}$ to $1.05\times10^7$.  The unbound companion mass
  caused by the SN impact as a function of time since explosion for
  each resolution is plotted in Figure~\ref{Fig:resolution}. The
  unbound mass is calculated by summing up the total mass of all
  particles that originally belonged to the He companion star but are
  unbound after the impact. In order to determine whether or not a
  particle is bound to the star, we calculated the total energy of
  each particle at each time step, $E_{\mathrm{tot}} =
  E_{\mathrm{kin}} + E_{\mathrm{pot}} + E_{\mathrm{in}}$, where
  $E_{\mathrm{kin}}$, $E_{\mathrm{pot}}$ and $E_{\mathrm{in}}$ are the
  kinetic energy (positive), the potential energy (negative) and the
  internal energy (positive), respectively. If $E_{\mathrm{tot}} > 0$,
  the particle is unbound.  Note that the center-of-mass motion of the
  star is subtracted when calculating the kinetic energy for each
  particle.

Figure~\ref{Fig:resolution} shows the amount of unbound
companion mass asymptotically approaches a final value at late
times. For the simulations that span a range of $400$ in mass
resolution from the lowest to the highest mass resolution the
stripped mass measured in those simulations deviates less
than $25\%$ from the highest resolution run. Therefore, our results
are clearly sufficiently well converged to allow a meaningful
comparison to observational constraints, which are still uncertain
by a factor of a few \citep{Leon07}. Note that we also carried out the 
convergence test for the amount of SN ejecta that are captured by 
the companion star after the SN explosion (for different resolutions of 
$\sim2.64\times10^4$--$1.06\times10^7$, the captured SN ejecta masses at the end of simulations are $0.0064, 0.0067, 
0.0075, 0.0072, 0.0060, 0.0059,$ and $0.0053\,\rm{M_{\odot}}$). We found that it is also 
sufficiently well converged. Therefore, we chose the level of 
5 million SPH particles to represent the He companion stars (which corresponds to
the total particles of $\sim10^7$) in all
following impact simulations.

\subsection{Typical Evolution after the SN~Ia Explosion}

Figure~\ref{Fig:rho} illustrates the temporal density evolution of the
SN ejecta and companion material of our hydrodynamics simulations for
the W7\_He01 model.  Before the SN explosion, the He companion star is
filling its Roche lobe. The WD explodes as a SN~Ia on the right side
of the companion star. The SN ejecta expand freely for a while before
hitting the surface of the donor star which faces towards the
explosion (see first snapshot). A shock wave develops while the
He-rich material is stripped-off from the companion star. This shock
wave propagates through the whole companion star and strips off
additional material from the far side of the companion. As the SN
ejecta flow around the companion star, a cone-shaped hole with an
opening angle with respect to the $x-$axis of ${\sim}35^{\circ}$ forms
in the SN ejecta (see Figure~\ref{Fig:rho}).  At the end of the
hydrodynamics simulations, about $0.027\,M_{\odot}$ of He-rich
material is stripped off due to the SN impact. The companion star
survives the explosion, but it is completely out of thermal
equilibrium and dramatically expanding due to the SN heating.
Compared with our previous work on WD+MS models \citep{Liu12}, this
effect is more significant since He companion stars have higher
orbital velocities.

Figure~\ref{Fig:rho_rot} shows how the orbital motion and the spin of
the He companion star affect the density structures of the SN ejecta
and the companion star. In this work, the hydrodynamics simulations
are carried out for He01 and He02 models (see Table~\ref{table:1}) by
including their orbital and spin velocities.  All simulated results
are shown in Table~\ref{table:1}. Note that ``W7'' means the W7
explosion model, the letter ``r'' indicates that the orbital motion
and spin of the companion star are included into the simulations. It
is evident that the additional unbound mass and kick velocity caused
by including the orbital motion and spin is very small (see
Table~\ref{table:1}), the difference being within $2\%$ compared to
non-rotating models.

\subsection{Parameter Study}

\begin{figure}
  \begin{center}
    \subfigure
    {\includegraphics[width=0.45\textwidth, angle=360]{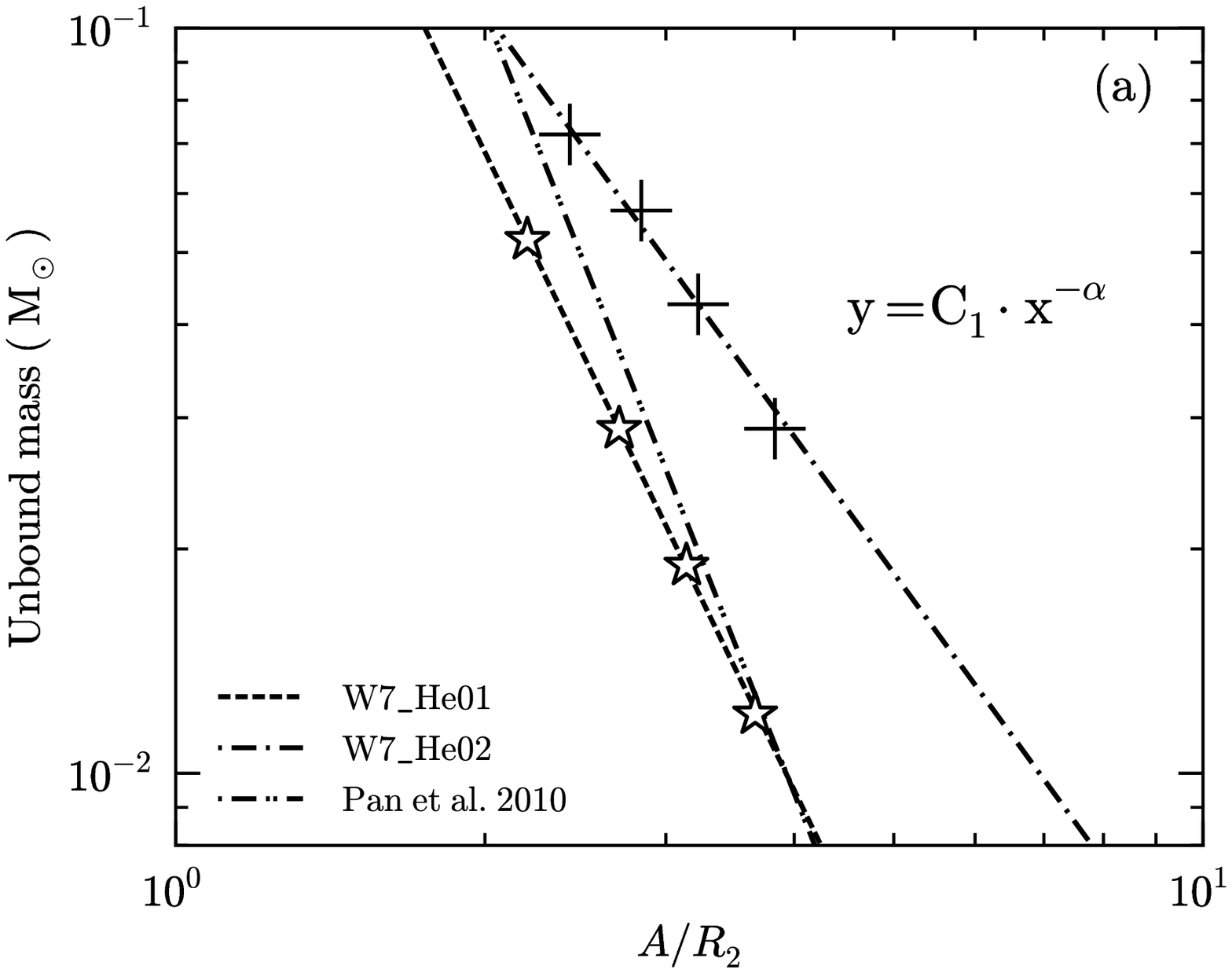}}
    \subfigure
    {\includegraphics[width=0.45\textwidth, angle=360]{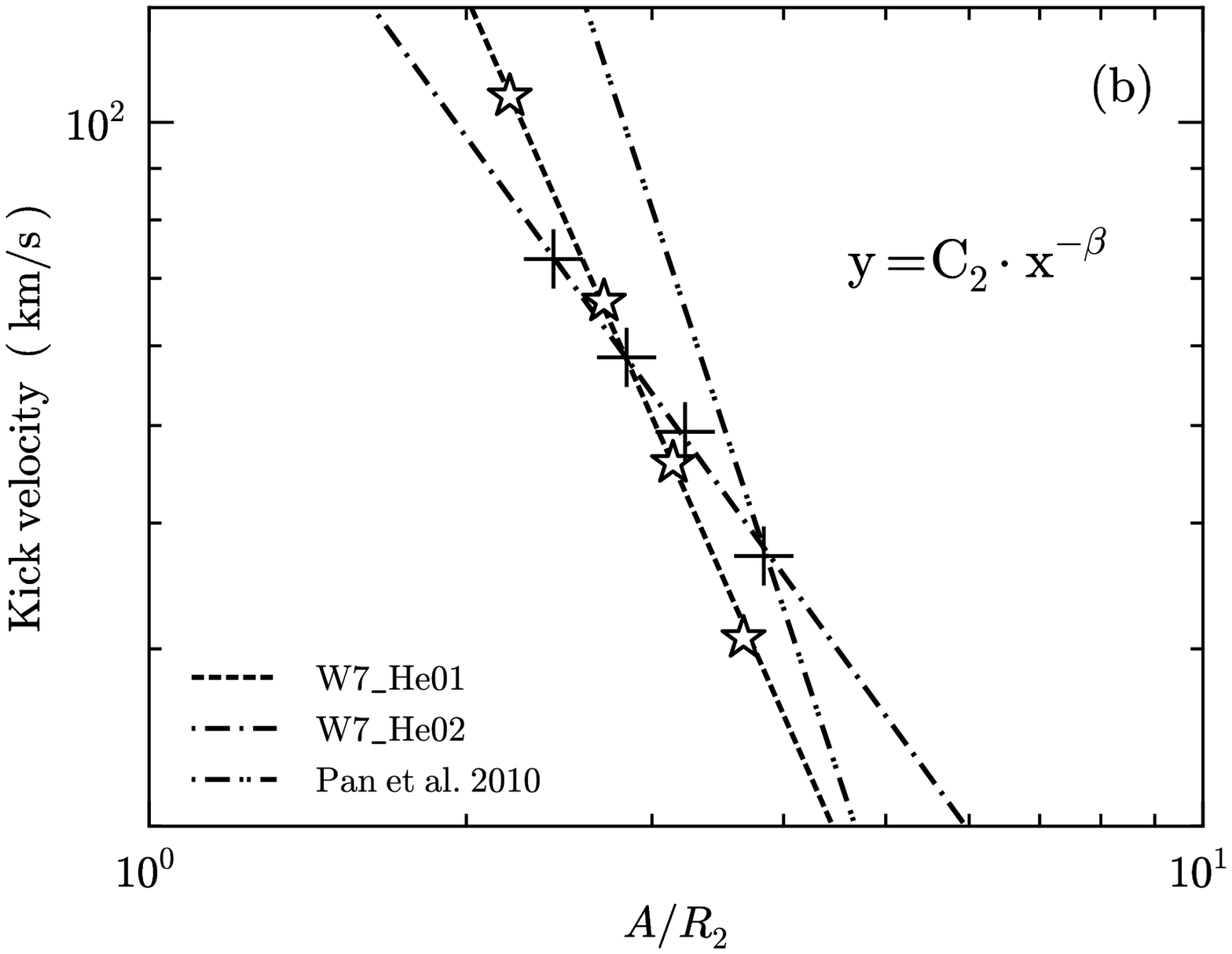}}
    \caption{Mass stripped-off from the companion (a) and kick
      velocity (b) as functions of the ratio of the orbital separation
      to the radius of the companion, $A/R_{2}$, for a given He
      companion model. The star and corss symbols represent the results of our impact simulations
      for the He01 model and He02 model. Lines show fitted power-law relations based on the
numerical simulation results.}
\label{Fig:fit}
  \end{center}
\end{figure}

At the end of the simulations, only 0.03--0.06\,$\mathrm{M}_{\odot}$
of He-rich companion material is found to be stripped off in impact
simulations of two different He companion models. Meanwhile, the
companion star receives a small kick velocity of
${\sim}$58--67\,$\mathrm{km\,s^{-1}}$ at the end of the
simulations. In order to explore the sensitivity of the numerical
results on the orbital separation, we run several simulations by
artificially adjusting the binary separations of the ``W7\_He01'' and
``W7\_He02'' models.  All other parameters are kept constant at the
values of the original model. Figure~\ref{Fig:fit} shows the unbound
mass and kick velocity as functions of the binary orbital separations,
which is consistent with other similar impact hydrodynamics
simulations \citep{Pan10, Pan12}.  For a given companion model, the
unbound mass decreases as the separation becomes larger. It is found
that this relation follows a power law in good approximation, and can
be fitted as (see Figure~\ref{Fig:fit}a):

   \begin{equation}
    \label{equation:1}
     M_{\mathrm{unbound}}= C_1  \left(\frac{A}{R_2}\right)^{-\alpha} \ \ \mathrm{M}_{\odot},
   \end{equation}

   where $A$ is the binary separation, $R_2$ is the radius of the He
   companion star at the onset of the SN explosion, $C_1$ is a
   constant and $\alpha$ is the power-law index. All fitting
   parameters are listed in Table~\ref{table:2}. Moreover, the
   dependence of the kick velocity, $v_{\mathrm{kick}}$, on $A/R_2$
   can also be fitted by a power law (see Figure~\ref{Fig:fit}b):

  \begin{equation}
    \label{equation:2}
     v_{\mathrm{kick}}= C_2  \left(\frac{A}{R_2}\right)^{-\beta} \ \ \mathrm{km\,s^{-1}},
   \end{equation} 

   where $C_2$ is a constant and $\beta$ is the power-law index (see
   Table~\ref{table:2}).

\begin{table} \renewcommand{\arraystretch}{1.5}
\begin{center}
  \caption{Fitting parameters for equation~(\ref{equation:1})
    and~(\ref{equation:2}) \label{table:2}}
\begin{tabular}{lcccc}
\tableline\tableline
  & \multicolumn{4}{c}{Fitting parameters} \\
Model & $\mathrm{C_{1}}$ & $\mathrm{\alpha}$ &  $\mathrm{C_{2}}$ & $\mathrm{\beta}$\\ 
\tableline
W7\_He01 & 0.54 & 2.96 & 689   & 2.37 \\
W7\_He02 & 0.34 & 1.75 & 247   & 1.38 \\
\tableline
\end{tabular}
\end{center}
\end{table}

The different companion star models lead to different fitting
parameters.  This indicates that the companion structure plays an
important role in our impact simulations also. For example, the binding
energy of the companion envelope would affect it. In order to compare
with other hydrodynamics simulations, the results of the He-WDc model
of \citet{Pan10} ($M_2 = 1.007\,M_{\odot}$, $A=4.0 \times
10^{10}\,\mathrm{cm}$ and $R_2=1.35 \times 10^{10}\,\mathrm{cm}$ at
the moment of the SN explosion)\footnote{The He-WDc model of
  \citet{Pan10}, was set up to mimic a system obtained from detailed
  binary evolution in WMCH09. This corresponds to our W7\_He02 model.}
is chosen to compare with our W7\_He02 model ($M_2 =
1.007\,M_{\odot}$, $A=7.04 \times 10^{10}\,\mathrm{cm}$ and $R_2=2.48
\times 10^{10}\,\mathrm{cm}$). The unbound mass and kick velocity in
their He-WDc model are more sensitive to the orbital separation than
in our W7\_He02 model (see Figure~\ref{Fig:fit}). The difference might
be caused by different companion structures. In their 1D calculations,
the mass-transfer from the He companion star was modeled by adopting a
constant mass-loss rate to mimic the work of WMCH09 (see
\citealt{Pan10}).  The orbital separation at the moment of the SN
explosion was calculated using the formulation of \citet{Eggl83}. In
our consistent binary calculations, however, we trace the details of
the binary evolution by treating the mass-transfer as RLOF, which also
fixes the separation of the binary system at this moment.

\begin{figure}
  \begin{center}
    \subfigure
    {\includegraphics[width=0.45\textwidth, angle=360]{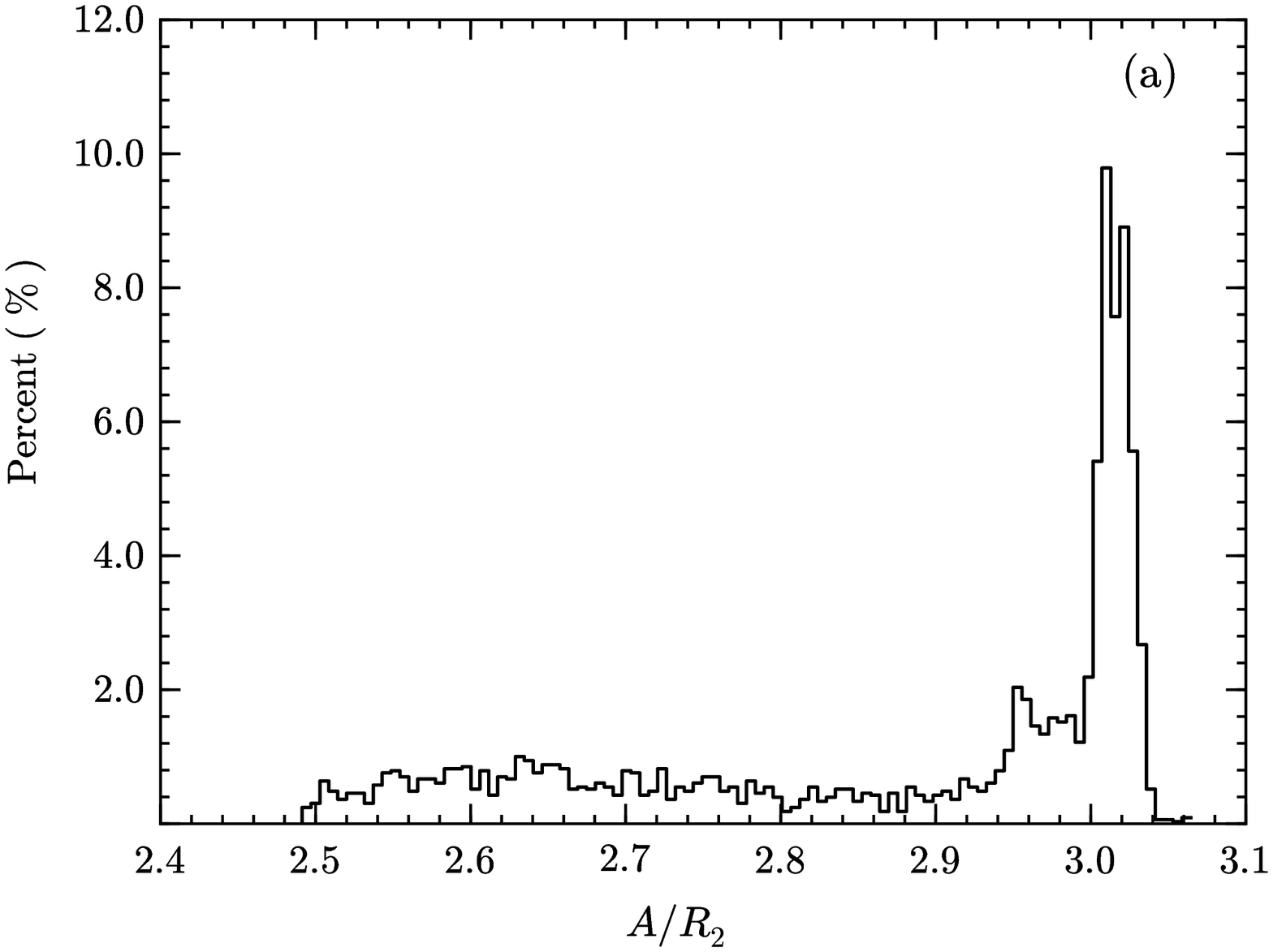}}
    \subfigure
    {\includegraphics[width=0.45\textwidth, angle=360]{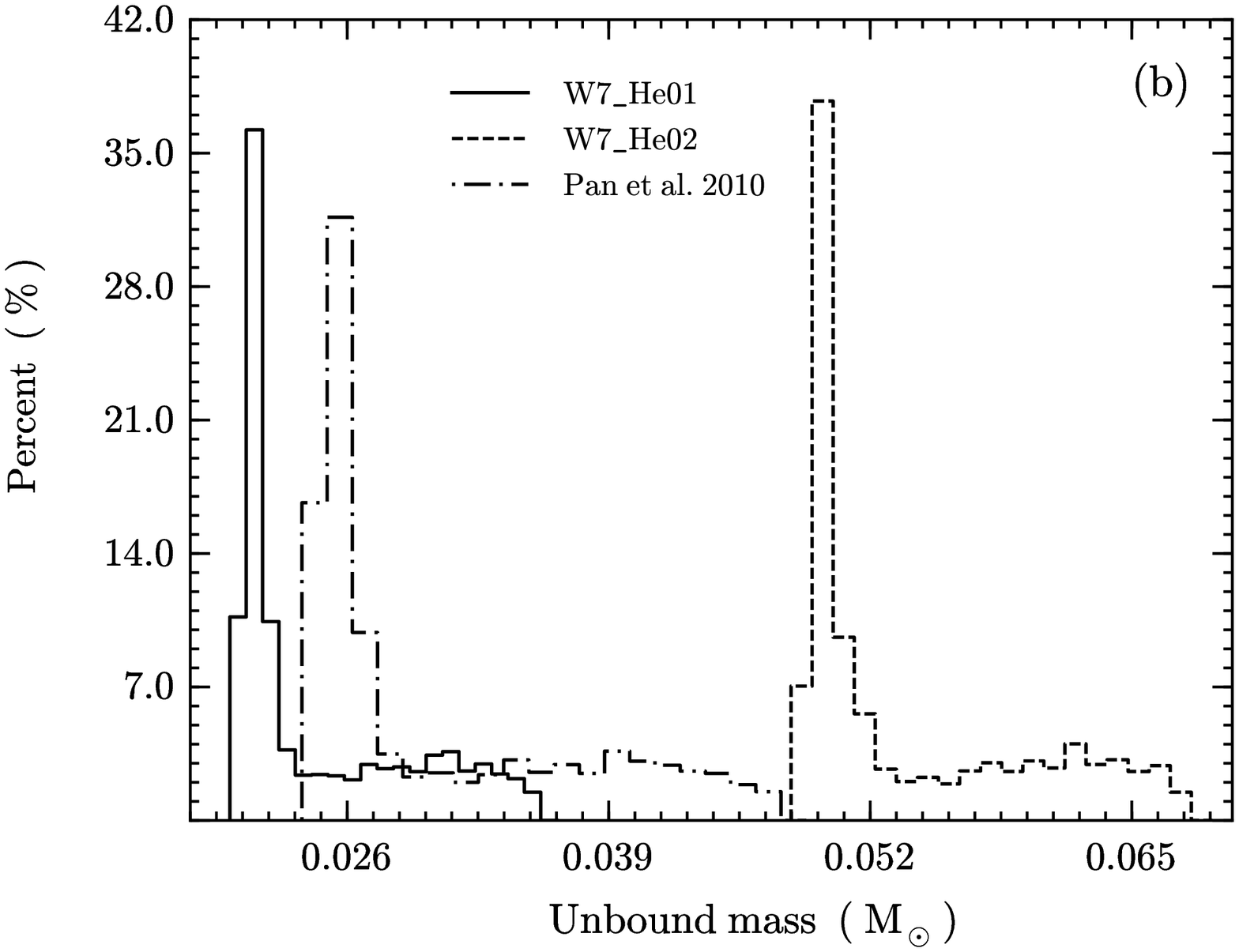}}
    \caption{ {\it Panel (a)\/}: Distribution of the parameter,
      $A/R_{2}$, based on the population synthesis results for WD+HE
      models \citep{Wang09b}.  {\it Panel (b)\/}: The corresponding
      distribution of unbound mass due to the SN impact. The unbound
      mass is calculated by using the power law relation of equation
      (3).}
\label{Fig:unboundmass}
  \end{center}
\end{figure}

Based on the distribution of the parameter $A/R_2$ in population
synthesis calculations of WMCH09 (see Figure~\ref{Fig:unboundmass}a),
we simply calculate the unbound masses due to the SN impact by using
equation~(\ref{equation:1}). The derived distribution for the unbound
mass is shown in Figure~\ref{Fig:unboundmass}b, where the peak unbound
mass ranges from $0.02\,\mathrm{M}_{\odot}$ to
$0.05\,\mathrm{M}_{\odot}$.\footnote{Note that we use the same 
power-law relation for different $A/R_2$ (different companion models) to 
predict the unbound masses. However, it is found that different companion models
would lead to different fitting parameters in SPH simulations (see 
Figure~\ref{Fig:fit}).}  The difference between the W7\_He02 and
He-WDc model, again indicates that the details of the companion
structures are important for the interaction of SN~Ia ejecta with the
companion star.

\section{DISCUSSION}
\label{sec:discussion}

\subsection{Effect of Ablation}
\label{sec:ablation}

After the SN explosion, the unbound He-rich material from the
companion star may result from two mechanisms: ablation (SN heating)
and stripping (momentum transfer). \citet{Pan12} found that the
stripped-to-ablated mass ratio for the He-WD scenario was about
0.5--0.8 in their impact simulations with the {\sc FLASH} code. They
argued that previous analytical or semi-analytical work underestimated
the unbound mass due to the neglect of ablation.

To obtain the stripped-to-ablated mass ratio, we compare the internal
energy of a companion particle to its kinetic energy once it becomes
unbound.  If the internal energy is larger (or smaller) than the
kinetic energy, we think the particle is ablated (or stripped). We
then use the SPH particle's ID to trace all these ablated (or
stripped) particles to the end of the simulations ($\rm{2000\,s}$
after the impact). The total ablated (stripped) mass is calculated by
summing the particles that are ablated (stripped) and still unbound at
the end of the simulations. Finally, we obtain a stripped-to-ablated
mass ratio of $\sim0.5$, which is consistent with the results of
\citet{Pan12}.

\begin{figure}
\centering
\includegraphics[width=0.6\textwidth, angle=360]{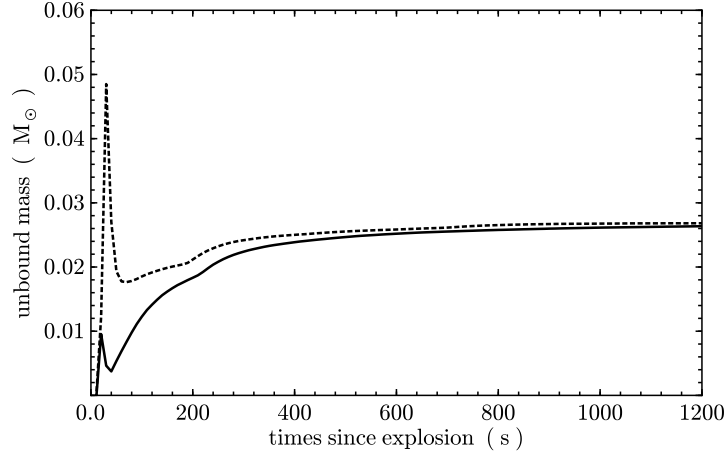}
\caption{ Unbound mass vs.\ simulation time in impact simulations with
  W7\_He01 model. The solid line shows the total mass of all particles
  with a total energy (kinetic plus potential energy) larger than
  zero. The dashed line also includes the internal energy in the sum.}
\label{Fig:ablation}
\end{figure}

Moreover, we calculate the amount of unbound mass by summing the total
mass of all unbound particles for each time step, where we do include
internal energy of the particle (i.e., $E_{\mathrm{tot}} =
E_{\mathrm{kin}} + E_{\mathrm{pot}} + E_{\mathrm{in}}$, which
corresponds to the dashed line in Figure~\ref{Fig:ablation}) or do not
(i.e., $E_{\mathrm{tot}} = E_{\mathrm{kin}} + E_{\mathrm{pot}}$, which
corresponds to the solid line in Figure~\ref{Fig:ablation}). The
companion particles are ablated and stripped and become unbound in
early stage of the explosion. As times goes by, the internal energy of
the particle converts into their kinetic energy. Moreover, some
ablated and stripped particles become bound again. Already $\rm{1000\,s}$ 
 after the explosion most of the internal energy deposited by the impact
 has been converted into kinetic energy (see Figure~\ref{Fig:ablation}).

\subsection{Hole in the Ejecta}

\begin{figure}
  \begin{center}
    \subfigure
    {\includegraphics[width=0.45\textwidth, angle=360]{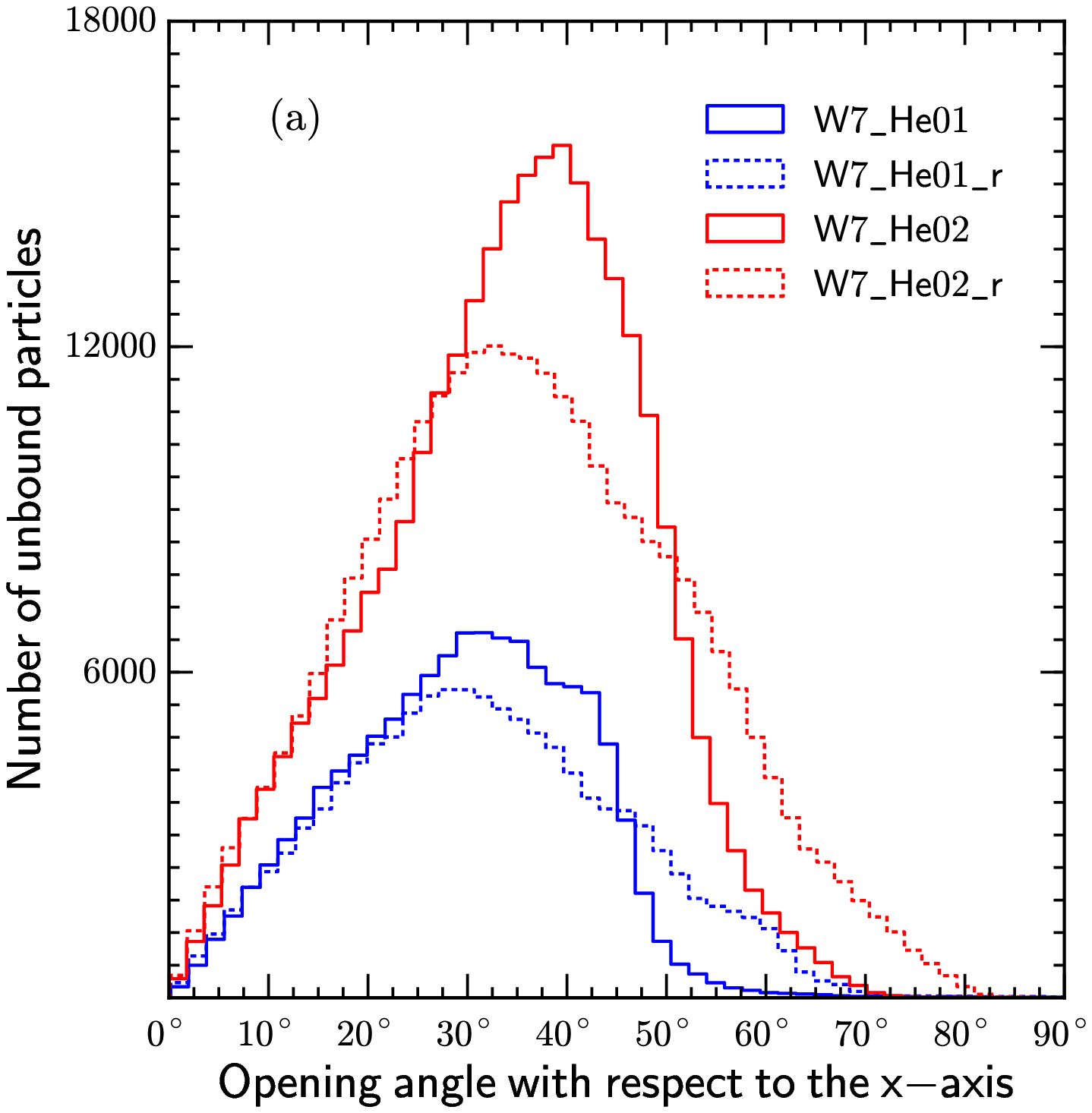}}
    \subfigure
    {\includegraphics[width=0.45\textwidth, angle=360]{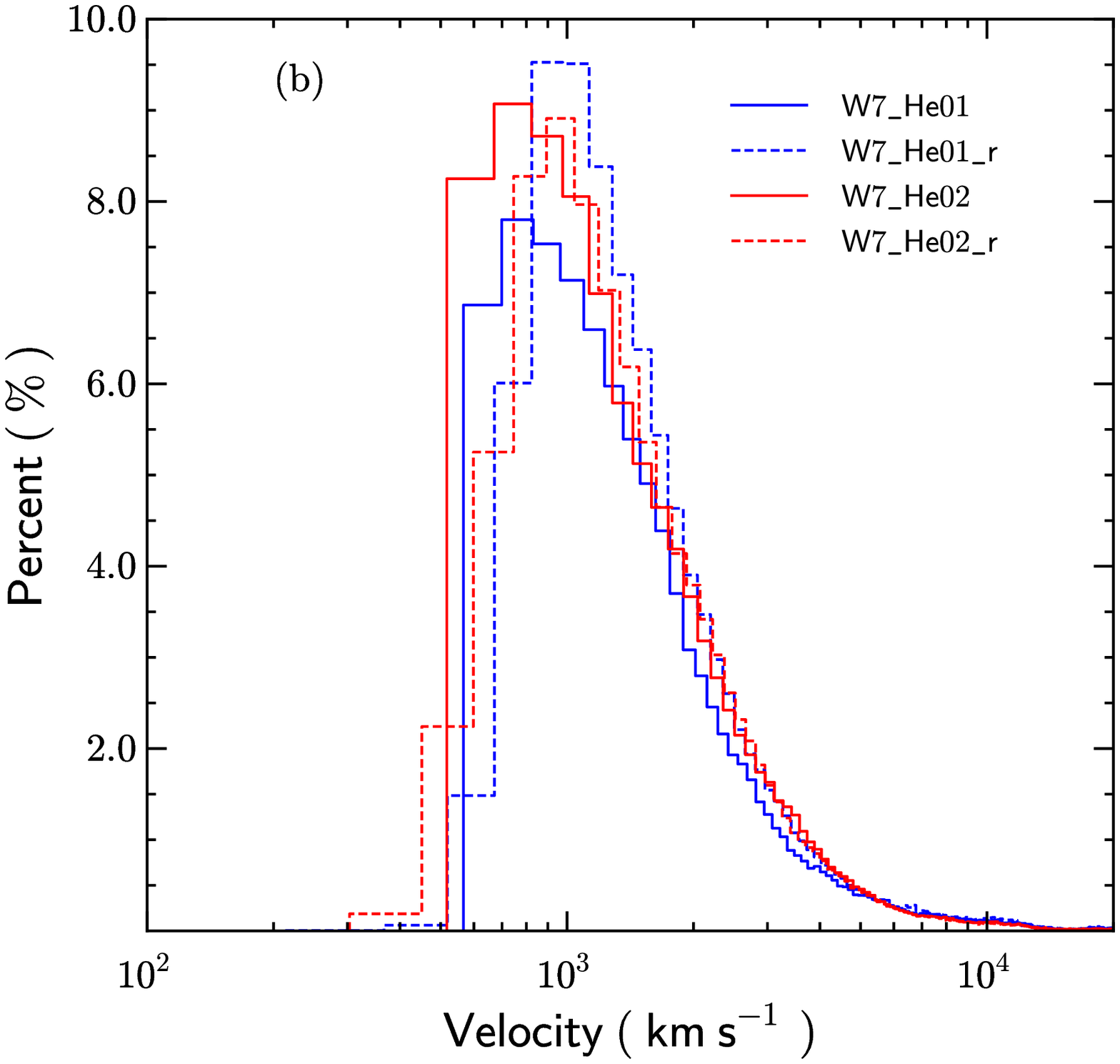}}
    \caption{ {\it Panel (a)\/}: distribution of opening angle of all unbound
      companion particles with respect to the $x-$axis at the end of
      the simulation. {\it Panel (b)\/}: velocity distribution of unbound
      companion material.} 
\label{Fig:vel_step}
  \end{center}
\end{figure}

The SN impact affects not only the companion star, but also the SN
ejecta themselves.  He-rich material is stripped off from the
companion due to the SN impact and largely confined to the downstream
region behind the companion star, creating a hole in the SN ejecta
with an opening angle of ${\sim}35^{\circ}$ with respect to the
$x-$axis in our simulation (see Figure~\ref{Fig:rho} and
Figure~\ref{Fig:vel_step}a). Recent hydrodynamic simulations suggest
that the cone-hole that is created during the interaction could remain
for hundreds of years \citep{Garc12}. \citet{Kase04} explored the
effect of a hole in the SN ejecta on spectra and light curves,
suggesting that the cone-hole might be a source of polarization of
SN~Ia spectra. For a recent review of spectropolarimetry measurements
of SNe~Ia see \citet{Wang08}.

\begin{figure}
  \begin{center}
    {\includegraphics[width=0.6\textwidth, angle=360]{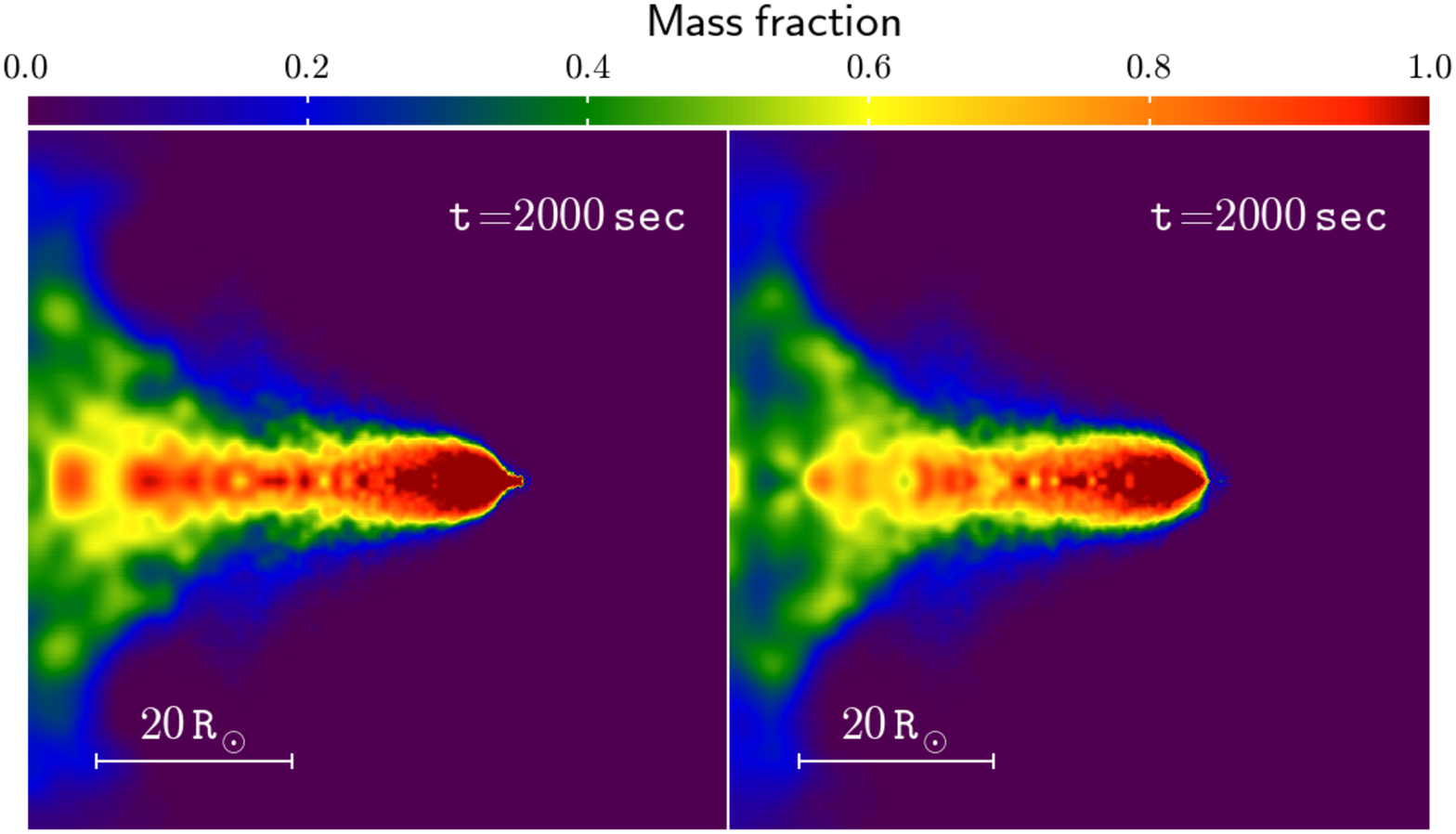}}
    \caption{Mass fraction of companion material to the SN ejecta in
      the hydrodynamics simulations of W7\_He01 ({\it left panel\/})
      and W7\_He01\_r model ({\it right panel\/}).  The blue end of
      the color table corresponds to pure SN ejecta material while a
      the red end of the color table represents pure companion
      material.}
\label{Fig:mix}
  \end{center}
\end{figure}

After the impact, stripped He-rich material is mixed with the SN
ejecta (see Figure~\ref{Fig:mix}). More SN ejecta material is found to
be mixed into the He-filled hole if the orbital motion and spin of the
He companion star are considered. The post-impact velocity
distributions of the stripped companion material are shown in
Figure~\ref{Fig:vel_step}b.  The peak velocity of
${\sim}800\,\mathrm{km\,s^{-1}}$ moves rightwards to
${\sim}1000\,\mathrm{km\,s^{-1}}$ when the orbital and spin velocities
of the companion star are included.  However, this peak velocity is
still smaller than the typical ejecta velocity of
${\sim}10^{4}\,\mathrm{km\,s^{-1}}$, which indicates that the stripped
He-rich material is largely hidden in the SN ejecta.  It might be
possible to detect it in late-time spectra of the SN when the ejecta
become transparent (see also \citealt{Pan12}). The high excitation
energy of He, however, may prevent the formation of He lines in the
nebular spectra.

\subsection{Accumulation of Ejecta on the Companion Star}
\label{sec:accumulation}

\subsubsection{Initial velocities of accreted SN ejecta}

The envelope of the companion star may be enriched by heavy elements
of the SN~Ia ejecta while its He-rich material is stripped off by the
SN impact. As a consequence a surviving companion star may show
unusual chemical signatures if a significant amount of SN ejecta
material is accumulated onto the donor star. \citet{Gonz09} concluded
that Tycho G has an unusually high nickel abundance, and they claimed
that it can be explained by the accumulation of SN ejecta. However,
the measured [Ni/Fe] ratio from a more recent study of \citet{Kerz12}
seems to be not unusual with respect to field stars with the same
metallicity.  Unusual abundances become a potential approach to
identify the He companion stars in SNRs after the nickel radioactively
decays.

\begin{figure}
  \begin{center}
    {\includegraphics[width=0.6\textwidth, angle=360]{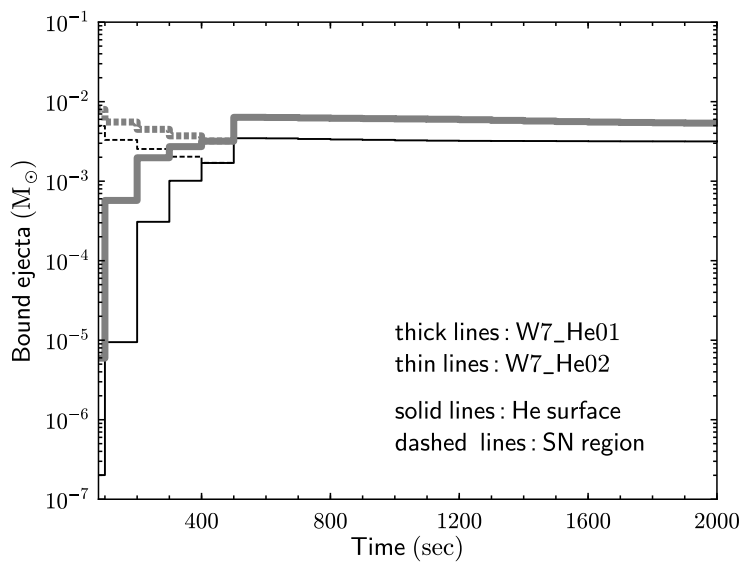}}
    \caption{Temporal evolution of bound ejecta masses. Early in the
      SN explosion, most of bound ejecta are found at the regions
      close to the SN explosion center.}
\label{Fig:bound_ejecta}
  \end{center}
\end{figure}

In the hydrodynamics simulations, we trace all bound particles that
originally belonged to the SN ejecta after the
explosion. Figure~\ref{Fig:bound_ejecta} illustrates the temporal
evolution of the amount of bound ejecta in the W7\_He01 and W7\_He02
models.  After the SN explosion, it takes some time for the ejecta
material to settle onto the surface of the companion star. Early after
the SN explosion, most of the bound ejecta material is found at
regions close to the SN explosion center. About $600-700\,\mathrm{s}$
after the impact, almost all bound ejecta particles fall onto the
companion (see figure~\ref{Fig:bound_ejecta}) and mix with the outer
layers of the star.  At the end of the simulations
(${\sim}2000\,\rm{s}$), the total amount of accreted SN ejecta is
$M_{\mathrm{tot}}{\sim}$3--5$\times 10^{-3}\,\mathrm{M_{\odot}}$
($M_{\mathrm{Ni}}{\sim}$0.8--1.6$\times 10^{-3}\,\mathrm{M_{\odot}}$
and $M_{\mathrm{Fe}}{\sim}$2--4$\times 10^{-3}\,\mathrm{M_{\odot}}$)
for the W7\_He01 and W7\_He02 models. The bound nickel mass is similar
to the results of the hydrodynamics simulations of \citet{Pan12}. In
order to check whether some bound ejecta particles become unbound
again at late times, we keep running the W7\_He01 and W7\_He02 models
until $7000\,\mathrm{s}$ after the impact. It is found that some bound
ejecta particles leave the companion star again, however, the change
is only 1\%--3\%. Therefore, we run all other simulations in this work
to only 2000--3000$\,\mathrm{s}$ to save computational resources.

\begin{figure}
\centering
\includegraphics[width=0.6\textwidth, angle=360]{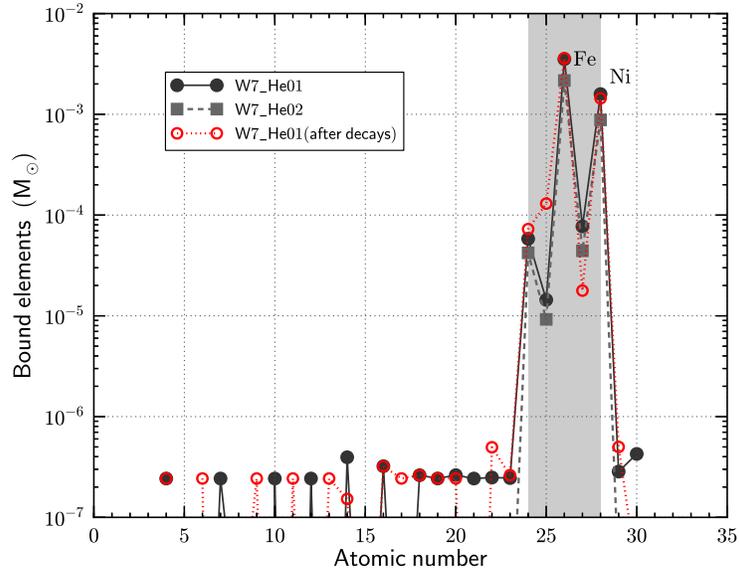}
\caption{Chemical composition of the accreted ejecta material
  ${\sim}2000\,\mathrm{s}$ after the explosion for the W7\_He01 and
  W7\_He02 models. The corresponding composition of the accreted ejecta after decays of unstable 
  isotopes for W7\_He01 model is shown with open circles.}
\label{Fig:elements}
\end{figure}

\begin{figure}
  \begin{center}
    {\includegraphics[width=0.45\textwidth, angle=360]{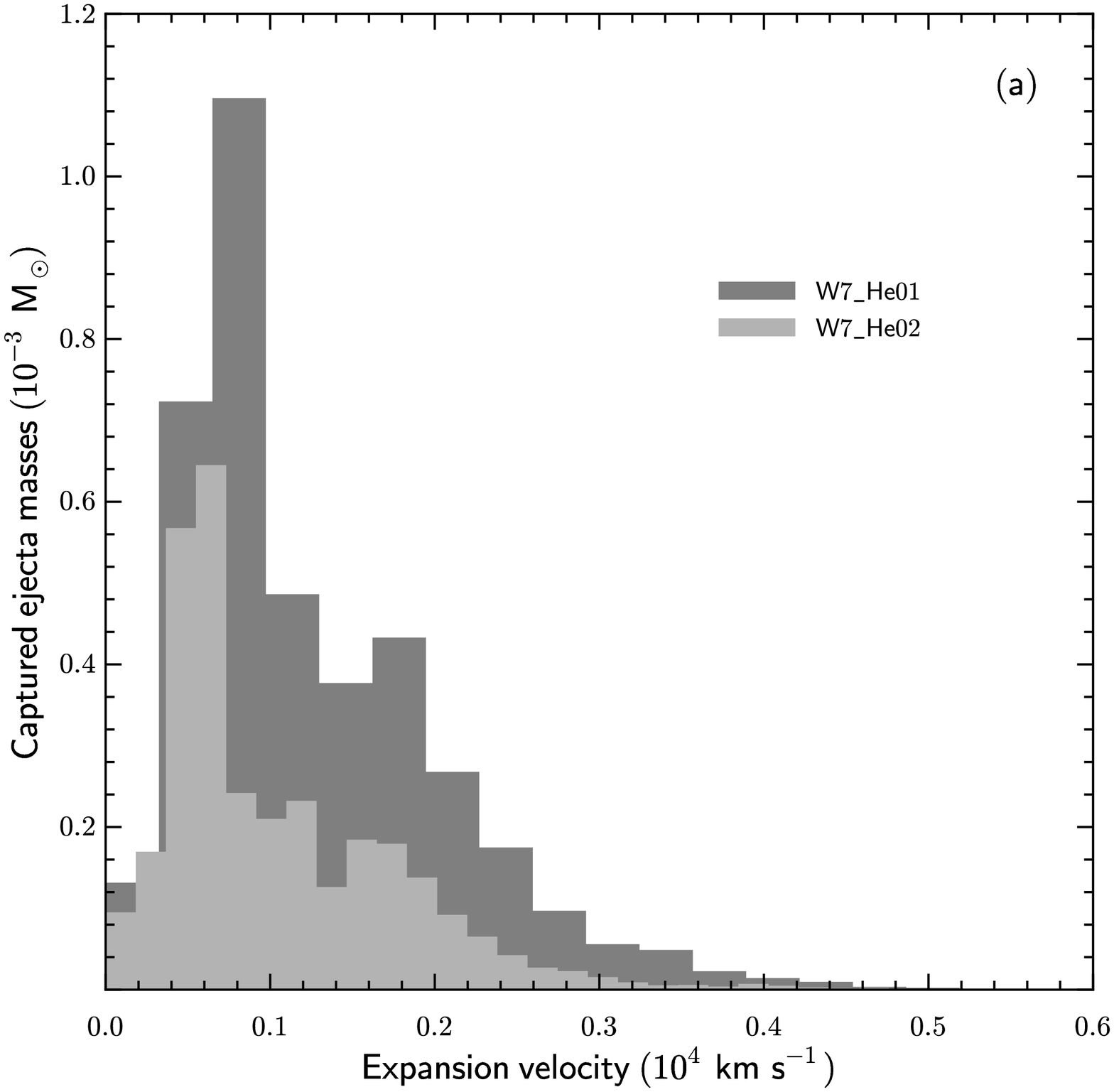}}
    {\includegraphics[width=0.45\textwidth, angle=360]{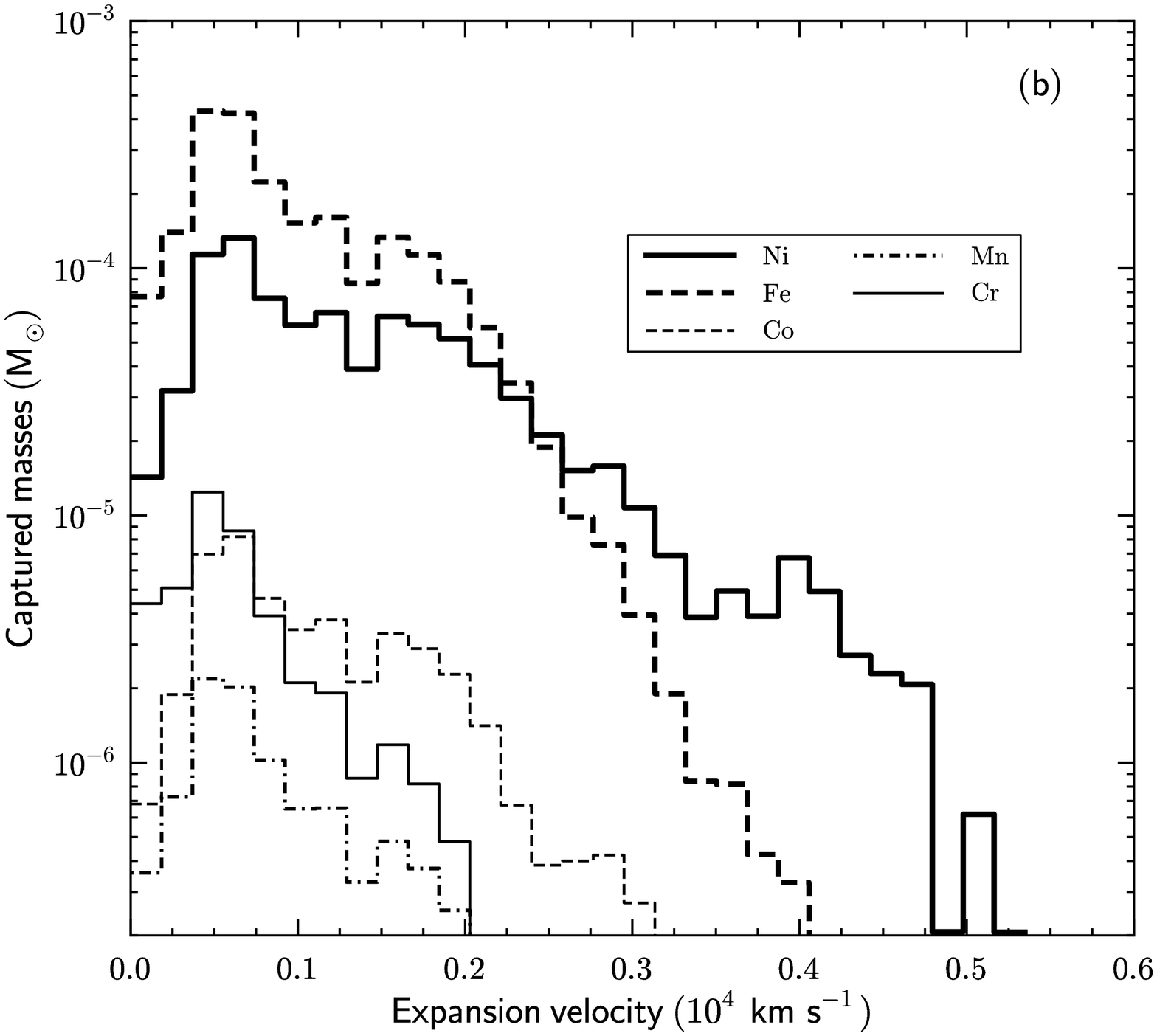}}
    \caption{{\it Panel (a)\/}: distribution of bound ejecta in their
      initial expansion velocity space (at $10\,\mathrm{s}$ after the
      explosion).  {\it Panel (b)\/}: similar to {\it Panel (a)\/},
      but only includes the iron-peak elements of the W7\_He02 model
      (Cr, Mn, Fe, Co and Ni) which correspond to the vertical gray
      range of Figure~\ref{Fig:elements}.}
\label{Fig:SN_r}
  \end{center}
\end{figure}

Figure~\ref{Fig:elements} shows the abundances of various chemical
elements accumulated from the SN ejecta onto the surface of the He
companion star at the end of the simulations. Iron-peak elements
(especially Fe and Ni) dominate the accreted ejecta (see the vertical
gray color range of Figure~\ref{Fig:elements}). Note that the masses
of unstable isotopes, such as $\leftidx{^{56}}{\mathrm{Ni}}$,
$\leftidx{^{57}}{\mathrm{Ni}}$ and $\leftidx{^{56}}{\mathrm{Co}}$ are
also included when summing up the Ni and Co masses.  In order to check
the original expansion velocity distribution of all accreted ejecta
particles, we trace the original positions of all bound ejecta
particles in the W7 model ($\mathrm{t=10\,s}$) based on their SPH ID
number. The result is shown in Figure~\ref{Fig:SN_r}a. Most of the
contamination is attributed to particles with low expansion velocity
in the SN ejecta (i.e., the innermost region of the W7 model).  The
typical peak expansion velocity of accreted ejecta material is
${\sim}10^{3}\,\mathrm{km\,s^{-1}}$.  This result can be explained by
the lower kinetic energy of those particles which makes it easier to
stay at the surface of the companion star after the momentum
transfer. The distribution of initial expansion velocities of all
accreted iron-peak elements (Cr, Mn, Fe, Co and Ni, which corresponds
to the vertical gray range of Figure~\ref{Fig:elements}) is further
shown in Figure~\ref{Fig:SN_r}b. Again, most accreted iron-peak
elements come from the low-velocity tail of the SN ejecta.
Therefore, we argue that the composition of the ejecta material that 
pollutes the companion star is very sensitive to the nuclear
burning at the center of the explosion and could, if detected, possibly
be used as a diagnostic of the explosion mechanism.

\subsubsection{Influence of orbital separation}

\begin{figure}
  \begin{center}
    {\includegraphics[width=0.6\textwidth, angle=360]{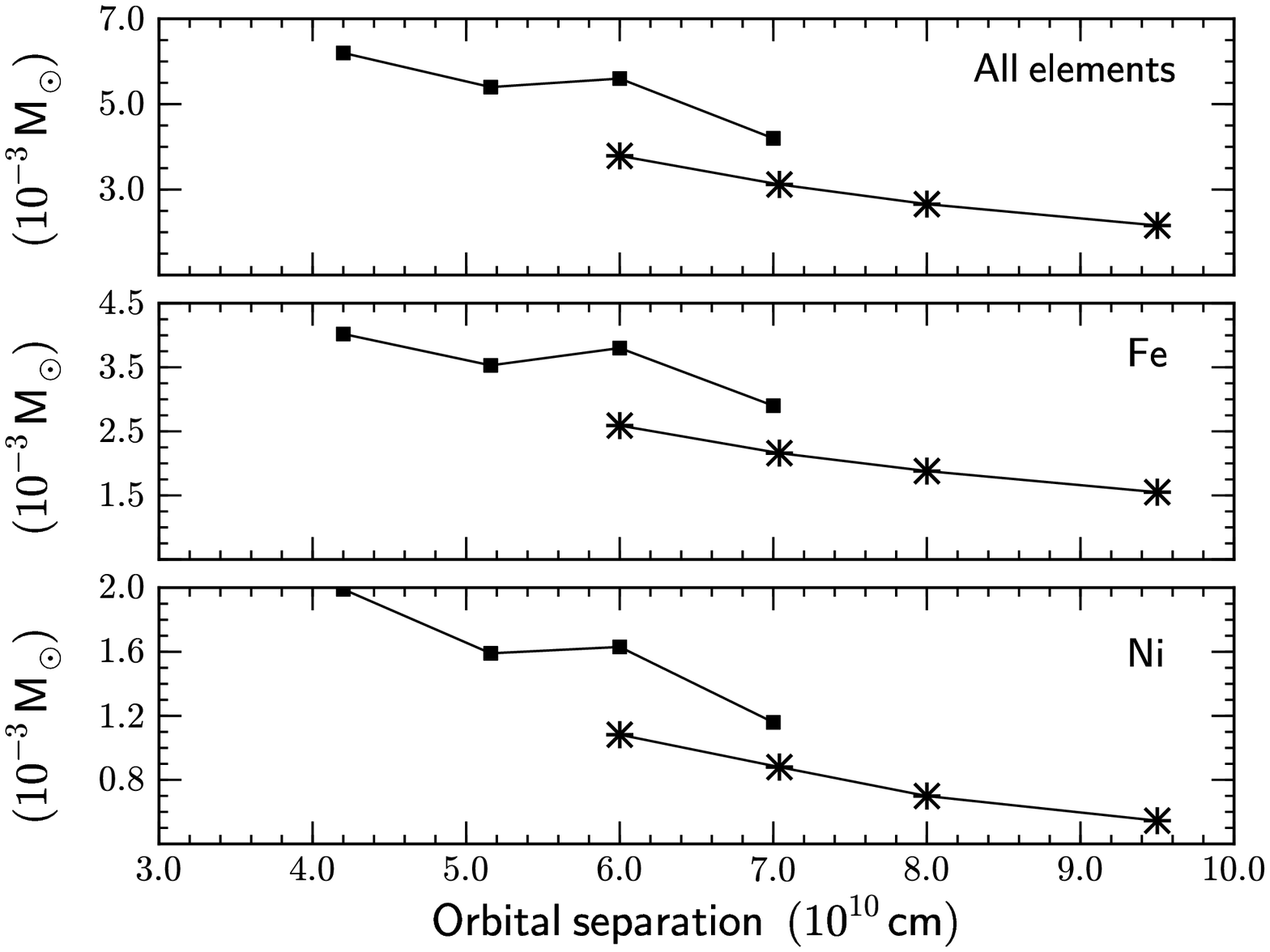}}
    \caption{ Dependence of bound-ejecta mass on the orbital
      separation in impact simulations with W7\_He01 ({\it square
        markers\/}) and W7\_He02 model ({\it star markers\/}). All values are measured 
       at the end of the SPH impact simulations.}
\label{Fig:contamination}
  \end{center}
\end{figure}

We checked the sensitivity of the level of total contamination with the
orbital separation for a given companion model. The orbital separation
is adjusted to cover the range of the $A/R_{2}$ parameter suggested by
population synthesis calculations as shown in
Figure~\ref{Fig:unboundmass}a.  Figure~\ref{Fig:contamination}
illustrates how the contamination depends on the orbital separation in
the W7\_He01 and W7\_He02 models.  The amount of SN~Ia ejecta
deposited on the surface of the He companion star is seen to vary with
the orbital separation for a fixed companion model. Larger orbital
separation leads to a lower ram pressure and also a smaller cross
section, reducing the contamination from SN~Ia ejecta. Note, however,
that the changes in the orbital separation of the W7\_He01 or W7\_He02
model are purely artificial.  Therefore, the effect of the nature of
the He companion star is ignored. The different amount of
contamination between W7\_He01 and W7\_He02 indicates that the details
of the companion structure are also important.

Moreover, the comparison between the results of the W7\_He01 (or
W7\_He02) and W7\_He01\_r (or W7\_He02\_r) models shows that the
asymmetry due to the orbital motion and spin of the He companion star
does not significantly affect the amount of the contamination of SN
ejecta in our hydrodynamical simulations (see Table~\ref{table:1}).

\subsubsection{Influence of the Explosion Energy}
 \label{sec:energy}
 
 A 1D parametrized pure deflagration of a $M_{\rm{Ch}}$ CO WD with a kinetic energy
 of $1.23\times10^{51}\,\rm{erg}$ (i.e., the W7 model, see
 \citealt{Nomo84}) is used to represent the SN~Ia explosion in our
 hydrodynamics simulations. However, different deflagration and
 detonation cases cover a typical range of kinetic energies of
 0.8--1.6$\times 10^{51}\,\mathrm{erg}$ \citep{Roep07, Game05, Seit12}. Here,
 we study how different explosion energies affect the interaction with
 the companion star.

For this purpose, we use the same method as \citet{Pakm08} to
artificially adjust the kinetic energy of the SN ejecta
$E^{i}_{\mathrm{kin,SN}}$ by scaling the velocities $v^{i}$ of the SN
particles based on the original W7 model (see also \citealt{Pakm08}):
 
\begin{equation}
  \label{equation:3}
  v^{i}= \sqrt{\frac{E^{i}_{\mathrm{kin,SN}}}{E^{\mathrm{W7}}_{\mathrm{kin,SN}}}}  v^{\mathrm{W7}},
\end{equation}

where $E^{W7}_{\mathrm{kin,SN}}$ and $v^{\mathrm{W7}}$ are the kinetic
energy ($1.23\times 10^{51}\,\mathrm{erg}$) and homologous expansion 
velocity of the ejecta (which corresponds to velocities of expanding shells of SN ejecta) of the original W7 explosion model. 
This scaling preserves the homologous expansion ($v \propto r$) of the
ejecta. Four additional W7-based models with different kinetic energies ($E^{i}_{\mathrm{kin,SN}} = 0.8, 1.0, 1.4$, and $1.6
\times 10^{51}\,\mathrm{erg}$) are studied. The lowest of these
kinetic energies is consistent with simulations of pure deflagrations
in CO WDs (e.g., \citealt{Roep07}). The upper limit is calculated by
assuming that a $M_{\rm{Ch}}$ WD consisting of an equal-by-mass
mixture of carbon and oxygen burns completely to
$\leftidx{^{56}}{\mathrm{Ni}}$.

\begin{figure}
  \begin{center}
    {\includegraphics[width=0.45\textwidth, angle=360]{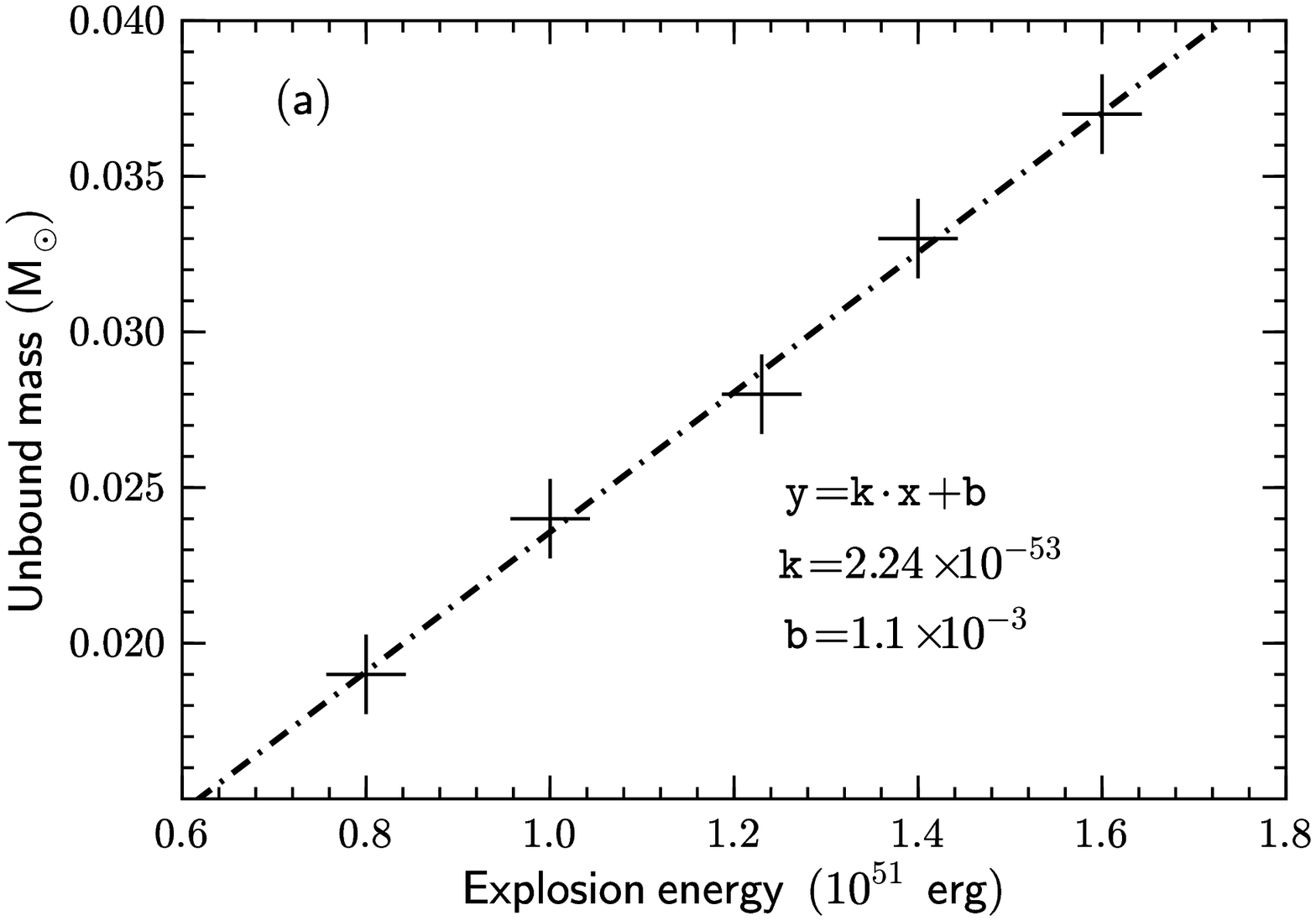}}
    {\includegraphics[width=0.45\textwidth, angle=360]{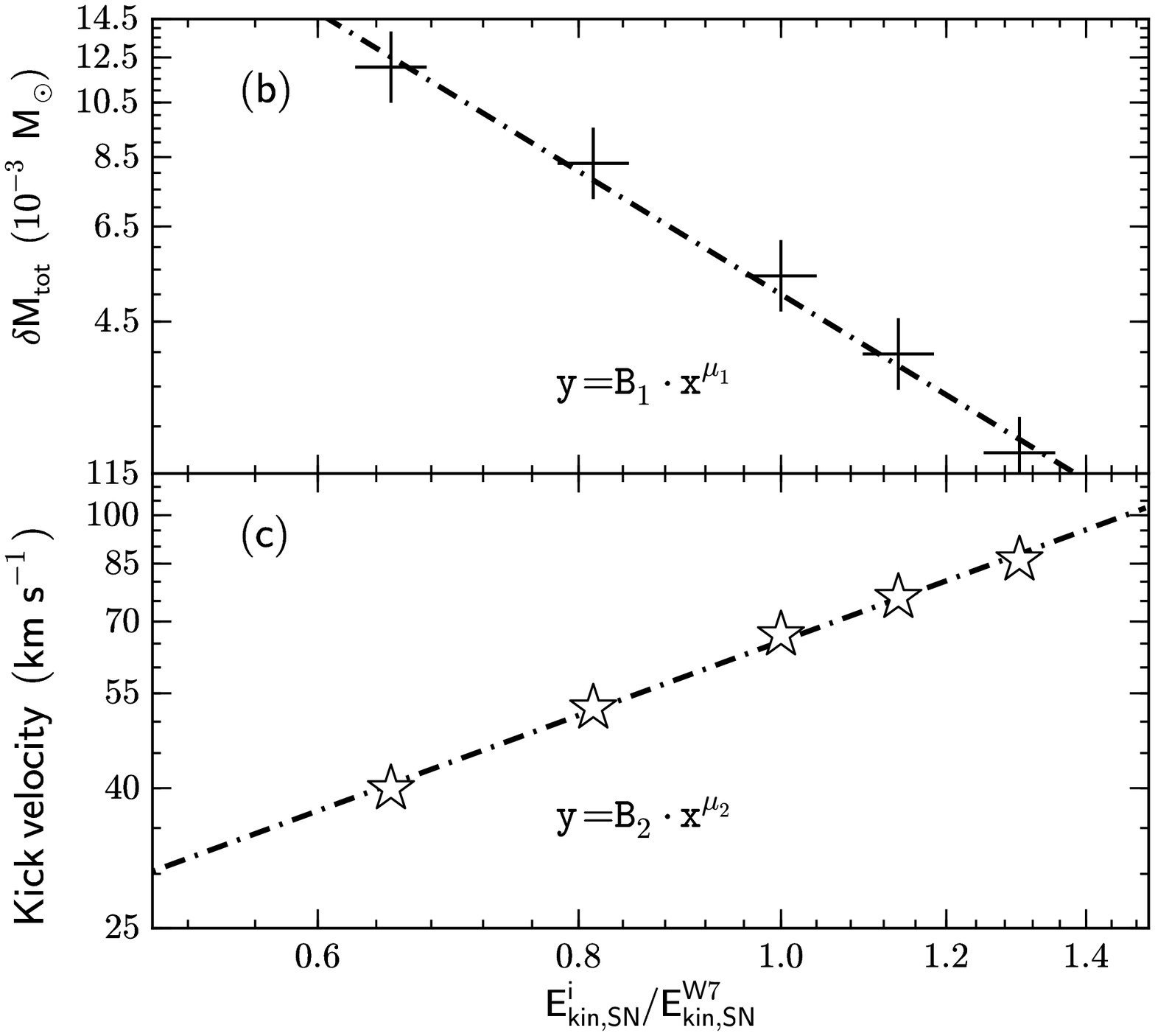}}
\caption{ {\it Panel (a)\/}: Total stripped mass as a function of the SN explosion energy 
          (see Table~\ref{table:1}). {\it Panel (b)\/}: 
          Power-law fit of the dependence of total accreted ejecta masses
          ($\mathrm{B_1=0.005,\ \mu_1=-2.133}$) on  the SN explosion energy  (Table~\ref{table:1}). 
          {\it Panel (c)\/}: Similar as
          {\it Panel (b)\/}, but for the kick velocity ($\mathrm{B_2=65.49,\ \mu_2=1.114}$).
          The star 
          and corss symbols represent the results of the impact simulations. 
          Lines show fitted linear (power-law) relations based on the numerical simulation results.}
\label{Fig:mass}
  \end{center}
\end{figure}

Using the ``He01\_r'' model as a representative case, we investigate
the influence of the SN explosion energy on the stripped companion mass, kick
velocity and deposited ejecta mass.  Numerical results for all
W7-based models are shown in Table~\ref{table:1}.  The stripped mass
increases linearly with SN explosion energy (see
Figure~\ref{Fig:mass}a). With a typical range of explosion energies of
0.8--1.6$\times10^{51}\,\rm{erg}$, the stripped companion mass by the SN impact changes
by a factor of two. This is consistent with the study of
\citet{Pakm08} for MS companion stars.  Moreover, the dependence of
kick velocity and captured ejecta mass on the explosion energy can be
fitted with a power law in good approximation (see
Figure~\ref{Fig:mass}b and Figure~\ref{Fig:mass}c).  It is not
surprising to find that the total contamination increases with
decreasing explosion energy. For smaller explosion energies, smaller
fractions of the ejecta are able to overcome the gravitational
potential energy at the end of momentum exchange. Therefore, a high
contamination of ${\sim}1.2 \times 10^{-2}\,\mathrm{M}_{\odot}$ is
found in impact simulations with the lowest explosion energy of $0.8
\times 10^{51}\,\mathrm{erg}$ (W708\_He01\_r model, see
Table~\ref{table:1}).

\subsubsection{Decay of unstable isotopes}
\label{sec:decay}

At the end of our simulations, the envelope of the surviving companion
star is enriched by the heavy elements accreted from the
low-expansion-velocity tail of the SN ejecta. In order to investigate
whether the surviving companion stars would be expected to show
observational over-abundance signatures, we estimate the ratio of the
bound Ni (or Fe) mass to the envelope He mass of a surviving companion
(see Table~\ref{table:1}) adopting the method of \citet{Pan10}.  We
also assume uniform mixing of the contaminants in the envelope.  With
two different companion models, the ratio of accreted Ni mass to the
companion envelope He mass, $M_{\mathrm{Ni}}/M_{\mathrm{He}}$, is
${\sim}$2--4$\times 10^{-3}$, which corresponds to a value of
$(M_{\mathrm{Ni}}/M_{\mathrm{He}})/(M_{\mathrm{Ni}}/M_{\mathrm{H+He}})_{\odot}\approx$24--48.
At the same time
$M_{\mathrm{Fe}}/M_{\mathrm{He}}{\sim}$4--9$\times10^{-3}$, which
corresponds to
$(M_{\mathrm{Fe}}/M_{\mathrm{He}})/(M_{\mathrm{Fe}}/M_{\mathrm{H+He}})_{\odot}\approx$2--6.
Here, we use the solar composition of \citet{Lodd03} to obtain the
corresponding value of $(M_{\mathrm{Ni}}/M_{\mathrm{H+He}})_{\odot}$
to compare with our simulation values.

\begin{figure}
  \begin{center}
    {\includegraphics[width=0.6\textwidth, angle=360]{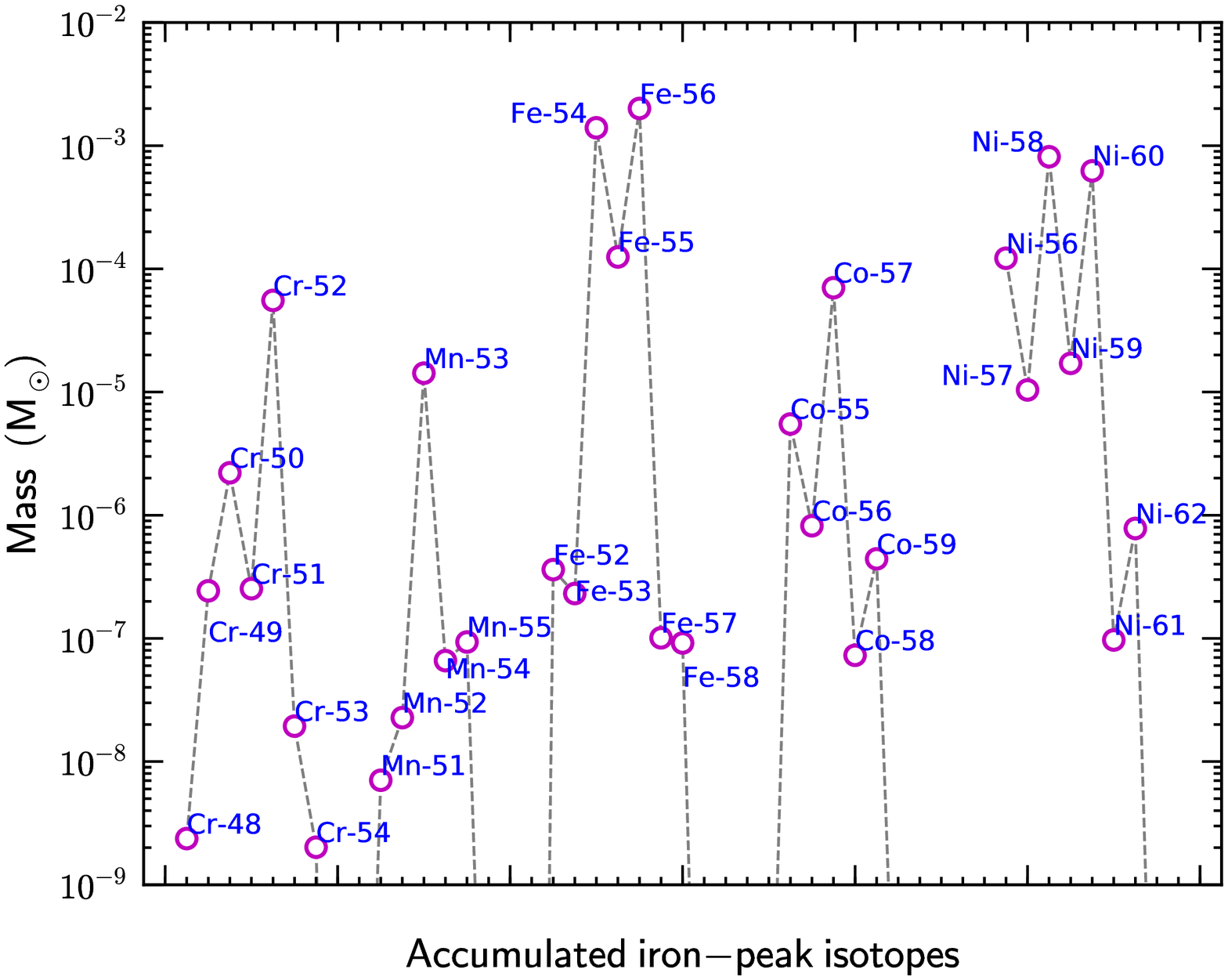}}
 \caption{Details of mass distributions of stable and unstable 
          isotopes of accreted iron-peak elements (Cr, Mn, Fe, Co and Ni, 
          i.e., vertical gray range of Figure~\ref{Fig:elements}) 
          in the impact simulations for W7\_He01 model.}
\label{Fig:unstable}
  \end{center}
\end{figure}

However, the above results neglect the radioactive decay of unstable
isotopes.  Figure~\ref{Fig:unstable} shows that the captured SN ejecta
material contains several unstable isotopes ($^{56}{\mathrm{Ni}}$,
$^{56}{\mathrm{Co}}$, $^{57}{\mathrm{Co}}$, $^{55}{\mathrm{Fe}}$,
etc), although stable species are the primary components.  The further
decay of unstable isotopes changes the long timescale Fe (or Ni)
abundances of the star. However, compared with the solar value of
iron/nickel-to-hydrogen plus helium of \citet{Lodd03}, our simulation
values after the radioactive decays are still larger (see Figure~\ref{Fig:elements}), providing a
possible way to identify a surviving companion star in SNRs by
detecting its unusual chemical abundance. We note that our previous
hydrodynamical simulations for WD+MS-like models showed that the
amount of contamination of SN ejecta is $\lesssim10^{-5}\,M_{\odot}$
(which corresponds to a small number of ejecta particles).  However,
this contamination of $\lesssim10^{-5}\,M_{\odot}$ is too small to
ensure whether it is a real contamination or not.

\begin{figure}
  \begin{center}
    {\includegraphics[width=0.45\textwidth, angle=360]{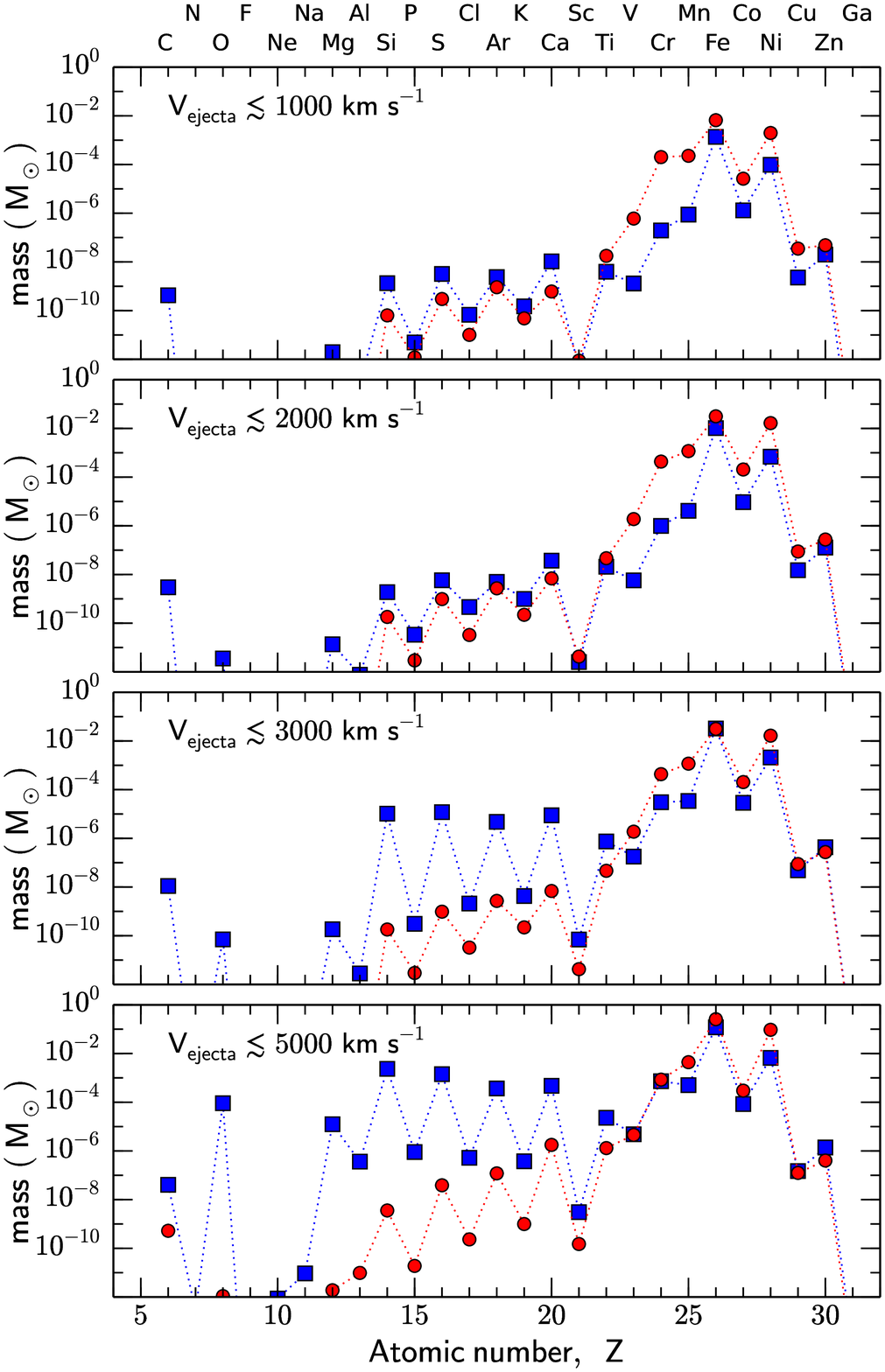}}
    \caption{Chemical composition of a delayed detonation model of `N100'
      ({\it squares\/}) \citep{Seit12} and the W7 model ({\it circles\/}) after radioactive decays of
      unstable isotopes. Different panels show details of the composition for
      different ejecta regions of the inner $1000, 2000, 3000,
      5000\,\mathrm{km\,s^{-1}}$. For example, the top (or bottom) panel shows the
      chemical composition of all ejecta material inside the spherical shell at ejecta velocity
      of $1000\,\mathrm{km\,s^{-1}}$ (or $5000\,\mathrm{km\,s^{-1}}$). } 
\label{Fig:models}
  \end{center}
\end{figure}

\begin{figure}
  \begin{center}
    {\includegraphics[width=0.45\textwidth, angle=360]{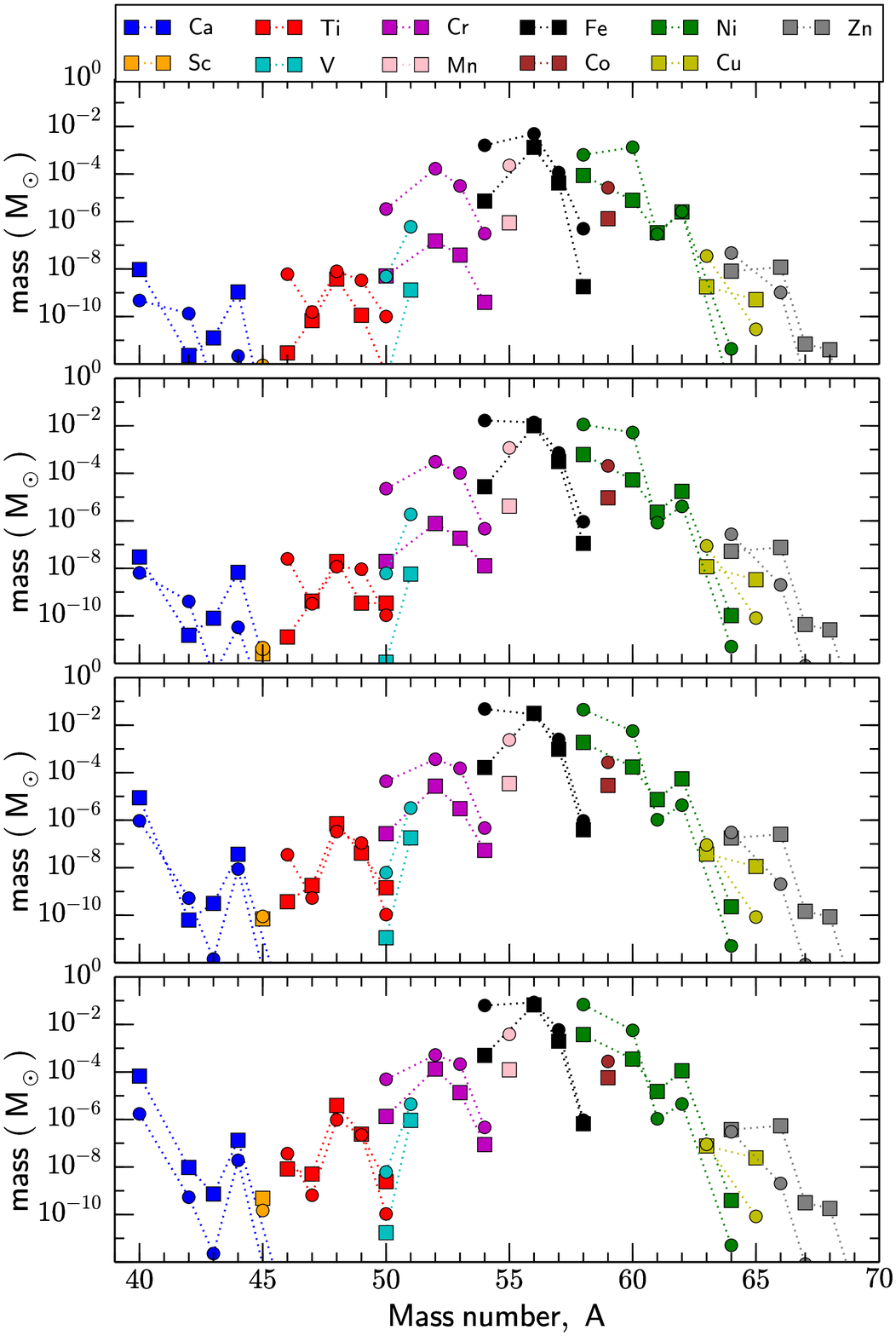}}
    \caption{Similar to Figure~\ref{Fig:models}, but for different
      isotopes of selected elements from $\mathrm{Ca}$ to
      $\mathrm{Zn}$. Squares are the delayed detonation model (i.e.,
      N100 model) and circles are the W7 model.}
\label{Fig:models1}
  \end{center}
\end{figure}

\subsubsection{Delayed detonation explosion model}
\label{sec:ddt}

Our hydrodynamics simulations with a classical W7 model show that a
surviving companion star in the WD+HE scenario can be significantly
enriched by heavy elements of the innermost SN ejecta. However, the
precise explosion mechanism of SNe~Ia remains unclear. Different
composition structures in various SN~Ia explosion models might affect
the abundance of captured heavy elements after the SN explosion.

To simply predict the effects of different explosion models, mass
distributions of ejecta elements of a delayed detonation model
\citep{Seit12} after radioactive decays of unstable isotopes are
compared with those of the W7 model. Here, we use the `N100 model' of
\citet{Seit12} to as an example realization of the delayed detonation
mechanism of an SN Ia (see also \citealt{Roep12}. The detailed comparisons
within different SN ejecta velocities are shown in
Figure~\ref{Fig:models} and Figure~\ref{Fig:models1}.  The delayed
detonation mechanism in the SD scenario has been suggested to be the most promising way
of producing observables in reasonable agreement with observations of
normal SNe~Ia \citep{Khok91}. In Section~\ref{sec:accumulation}, it is found that most of
the captured heavy elements come from the innermost SN ejecta (see
Figure~\ref{Fig:SN_r}). Therefore, we restrict the detailed
comparisons in Figure~\ref{Fig:models} and Figure~\ref{Fig:models1} to
SN ejecta regions with an expansion velocity of $\lesssim
5000\,\mathrm{km\,s^{-1}}$.

In Figure~\ref{Fig:models}, the inner ejecta of the `N100 model' show
distinctly smaller masses of stable Ni, Co, Mn and Cr compared to
those of the `W7 model', but basically similar stable Fe
mass. Therefore, we can roughly expect that a surviving companion star
may be less enhanced with Ni, Co, Mn and Cr due to the relatively
ineffective enrichment if we use the N100 model instead of the W7
model to carry out the same impact hydrodynamics simulations.
However, we do not expect the N100 model will lead to a surviving
companion star with significantly different Fe abundance compared to a
surviving companion impacted by W7 ejecta.  Moreover, a significant
amount of stable Si, Ca, S and Ar within $5000\,\mathrm{km\,s^{-1}}$
in the N100 model indicates that its surviving companion star might
show observable signature of Si, Ca, S and Ar enhancement (see
Figure~\ref{Fig:models}).

\subsection{Indicators of a surviving companion star}
\label{sec:observability}

In case some SNe~Ia originate from the WD+HE $M_{\rm{Ch}}$ scenario,
our simulations indicate that surviving companion stars would show
characteristic observational features due to contamination by SN
ejecta (see Section~\ref{sec:accumulation}). This may help to identify
a surviving companion star even a long time after the SN explosion.

In our simulations, it is found that the kick velocity received by a
companion is ${\sim}$58--67$\,\mathrm{km\,s^{-1}}$. WMCH09 showed that
the He companion has an orbital velocity of
${\sim}$300--500$\,\mathrm{km\,s^{-1}}$ at the moment of the SN
explosion. This indicates a surviving companion star will have a high
spatial velocity that is similar to its pre-explosion orbital
velocity.

\begin{figure}
  \begin{center}
    {\includegraphics[width=0.6\textwidth, angle=360]{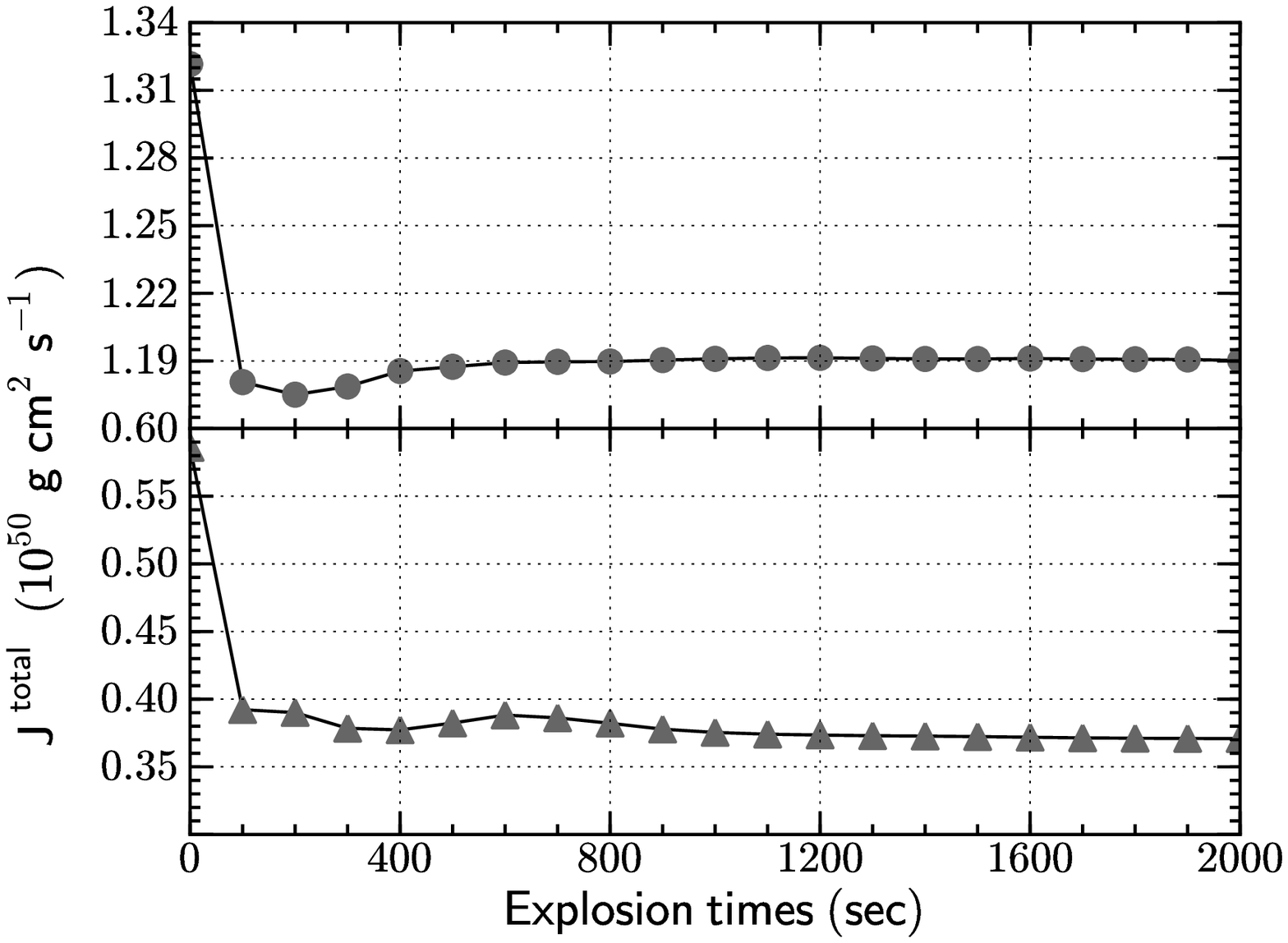}}
    \caption{ Temporal evolution of the total angular momentum of the
      companion star in W7\_He01\_r ({\it top panel\/}) and
      W7\_He02\_r ({\it bottom panel\/}) model (see
      Table~\ref{table:1}). }
\label{Fig:angular_t}
  \end{center}
\end{figure}

A small stripped mass (${\sim}$0.03--0.06$\,\mathrm{M}_{\odot}$) is
insufficient to remove the total angular momentum of a He companion
(only 13\%--38\% of initial angular momentum are lost from the star,
see Figure~\ref{Fig:angular_t}). Therefore, we expect that He
survivors would be rapid rotators (for a detailed discussions of
post-impact rotation of surviving companion stars, see \citealt{Liu13,
  Pan13}). In WMCH09, it was shown that the pre-explosion rotational
velocities of companion stars in the WD+HE $M_{\rm{Ch}}$ channel are
120--380$\,\mathrm{km\,s^{-1}}$ assuming that the rotation of the
companion star is phase-locked to its orbital motion due to tidal
forces.

At the end of our simulations, about
0.028--0.056$\,\mathrm{M}_{\odot}$ of He-rich material are stripped
off from the He companion stars. Full radiative transport calculations
with the results of our hydrodynamics simulations are required to
assess the possibility of detecting He lines in the nebular spectra of
the modeled events.
  
After the SN impact, a surviving companion star dramatically puffs up
due to the significant SN heating, and it would become a luminous He
star near the SNR center while its equilibrium is reestablished during
several centuries after the explosion \citep{Pan13}. Moreover, it may
be a rapidly rotating star with a high spatial velocity (see
\citealt{Pan13}).

One way to verify the WD+HE progenitor scenario is by identifying a
corresponding surviving star in a SN remnant.

\section{SUMMARY AND CONCLUSIONS}
\label{sec:conclusions}

The primary goal of this work has been to investigate the interaction
of SN~Ia ejecta with the companion star within the WD+HE $M_{\rm{Ch}}$
explosion scenario. We mainly focused on whether or not a surviving
companion shows an unusual abundance signature after the SN
explosion. We have performed 3D hydrodynamics impact simulations
employing the SPH code {\sc Stellar GADGET}. The effect of the orbital
motion and spin of the companion star were also taken into account. Two
representative He companion models were obtained from 1D consistent
binary evolution calculations with Eggleton's stellar evolution code,
treating the mass loss of the donor star as RLOF. Our main conclusions
are summarized as follows:

\begin{itemize}

\item In the WD+HE $M_{\rm{Ch}}$ scenario, it is found that only
  ${\sim}$2\%--5\% of the initial companion mass can be stripped off
  due to the SN impact. The star receives a small kick velocity of
  ${\sim}$58--67$\,\mathrm{km\,s^{-1}}$.
\item A power-law relation similar to that of \citep{Pan10, Pan12} is
  found between the unbound mass (or kick velocity) and the orbital
  separation for a given companion star model.
\item The orbital motion and spin of a He companion star do not
  significantly affect the amount of unbound mass and kick velocity
  caused by the SN impact.
\item Our simulations predict that a surviving companion star in the
  WD+HE $M_{\rm{Ch}}$ channel moves with a high spatial velocity and
  should be a fast rotator after the SN explosion.
\item The He companion star is enriched by the heavy elements with low
  expansion velocity of the SN~Ia ejecta. The total contamination is
  $\gtrsim 10^{-3}\,\mathrm{M}_{\odot}$, providing a potential way to
  identify a survivor after the SN explosion.
\item The amount of contamination from SN~Ia ejecta decreases with the
  increase of SN explosion energy and can be fitted with a power-law
  relation in good approximation.
\end{itemize}

Our results are based on the standard SN~Ia explosion model `W7'
\citep{Nomo84, Maed10}.  The comparison in Section~\ref{sec:ddt}
indicates that more comprehensive investigations with various
state-of-the-art explosion models are needed to reliably predict
whether the surviving companion star in the WD+HE $M_{\rm{Ch}}$
channel would show unusual abundances.

\section*{Acknowledgments}

We thank U.\ Noebauer for very useful discussions.  Z.W.L and Z.W.H
thank the financial support from the MPG-CAS Joint Doctoral Promotion
Program (DPP) and Max Planck Institute for Astrophysics (MPA).  This
work is supported by the National Basic Research Program of China
(Grant No. 2009CB824800), the National Natural Science Foundation of
China (Grant Nos. 11033008 and 11103072) and the Chinese Academy of
Sciences (Grant N0. KJCX2-YW-T24).  The work of F.K.R was supported by
Deutsche Forschungsgemeinschaft (DFG) via the Emmy Noether Program (RO
3676/1-1) and by the ARCHES prize of the German Federal Ministry of
Education and Research (BMBF). The work by K.M. is supported by World
Premier International Research Center Initiative (WPI Initiative),
MEXT, Japan, and by Grant-in-Aid for Scientific Research for Young
Scientists (23740141).  S.T. is supported by the DFG through the
Transregional Collaborative Research Centre `The Dark Universe'
(TRR~33). The simulations were carried out at the Computing 
Center of the Max Planck Society, Garching, Germany.

\bibliographystyle{apj}

\bibliography{ref}

\end{document}